\documentclass[]{memtensor}

\usepackage[toc,page,header]{appendix}
\usepackage[utf8]{inputenc}
\usepackage{amssymb}
\usepackage{longtable}
\usepackage{makecell}
\usepackage{array}
\usepackage{float}
\usepackage{threeparttable}
\graphicspath{{./}}

\title{SparseX: Efficient Segment-Level KV Cache Sharing for Interleaved LLM Serving}

\author[1]{Quqing Zhang}
\author[1]{Kai Chen}
\author[1]{Ning Liao}
\author[1]{Zehao Lin}
\author[]{Bo Tang}
\author[1]{Feiyu Xiong}
\author[1]{Zhiyu Li}
\author[\dagger]{Xiaoxing Wang}
\affiliation[1]{MemTensor (Shanghai) Technology Co., Ltd.}
\contribution[\dagger]{Corresponding author}

\abstract{
In long-context LLM serving, the prefill stage often dominates time-to-first-token and computational cost. Although Prefix Cache in vLLM/PagedAttention~\cite{kwon2023pagedattention} has been widely used to reuse identical prompt prefixes, repeated content in practical applications frequently appears as non-prefix, cross-request, cross-turn, and cross-agent segments, which makes conventional cache mechanisms insufficient. This paper presents \method, a segment-level \kv{} sharing method for common serving scenarios. \method{} uses contiguous token segments as reuse units and exploits \sparseq{} indices that naturally arise in \kv{} reuse workloads to estimate the key tokens that require correction. Based on this estimate, \method{} performs Sparse-KV Recomputation within a single forward pass, thereby restoring cross-segment contextual interactions under complex interleaved reuse patterns while avoiding additional models or separate preprocessing stages for token selection. \method{} further implements a \fullsparse{} hybrid attention mode based on a layer-specific threshold: early layers retain full attention to obtain a more stable token-importance signal, and later layers switch to sparse recomputation to improve reuse quality on complex long-context tasks. We implement SparseX-vLLM on top of vLLM, integrating segment-level cache lookup, PagedAttention management, RoPE~\cite{su2024roformer} alignment, \sparseq{} token selection, and FlashAttention~\cite{dao2022flashattention} backends into a unified execution path. \method{} is model-agnostic, training-free, and compatible with Prefix Cache, and it provides unified support for common online serving scenarios including multi-round chat, retrieval-augmented generation (RAG)~\cite{lewis2020rag}, and agent workflows.
}
\checkdata[Code]{\url{https://github.com/MemTensor/SparseX}}

\newcommand{\method}{SparseX}
\newcommand{\kv}{KV Cache}
\newcommand{\sparseq}{Sparse-Q}
\newcommand{\fullsparse}{full+sparse}

\begin{document}
\maketitle

\section{Introduction}

\subsection{Background}

In recent years, large language models (LLMs) based on the Transformer architecture~\cite{vaswani2017attention} have become core infrastructure for dialogue systems, retrieval-augmented generation (RAG), and complex agent workflows. In online serving, LLM inference usually consists of two stages. The prefill stage performs a forward pass over the input context and constructs the \kv{}, while the decode stage autoregressively generates tokens using the existing \kv{}. In typical long-context applications, prefill dominates end-to-end latency and compute cost because it must execute multi-layer self-attention and feed-forward computation over a context of length \(N\), while generating and storing \kv{} tensors at every layer. As context length, request concurrency, and interaction depth increase, the computation and storage overhead of the \kv{} grows further and becomes a major bottleneck in LLM serving.

In real systems, substantial repeated content often exists across different requests and even across different turns of the same session. Multi-round dialogue repeatedly uses historical messages; RAG workloads may repeatedly retrieve the same knowledge snippets for different user requests; and agent workflows may share intermediate reasoning materials, tool outputs, or candidate drafts across multiple agents. Reusing the \kv{} associated with these repeated segments during the prefill stage of a new request can avoid redundant forward computation over repeated tokens, which can substantially reduce latency and cost. However, arbitrary-position \kv{} reuse is not a simple memory-copy operation. Because self-attention couples a segment representation with its prefix context, the hidden states and \kv{} tensors of the same text segment often differ under different prefixes and different concatenation positions. Therefore, reuse in realistic settings is a context-dependent and position-dependent approximate computation problem: under an output-quality constraint, the system must determine which tokens can directly reuse cached \kv{} tensors, which tokens must be recomputed, and how this mechanism can be reliably integrated into a high-performance inference engine.

Recent work has advanced this problem along three directions. One line extends the hit range of Prefix Cache and attempts to directly reuse non-prefix segments. A second line corrects reuse error through selective \kv{} recomputation. A third line combines sparse attention or auxiliary models to predict which tokens should be recomputed. These methods usually rely on specific prompt structures, application scenarios, or additional computational components, which makes it difficult for them to simultaneously cover the complex interleaved reuse patterns that arise in multi-round chat, RAG, and agent workflows.

Motivated by these limitations, we study and propose \method, a general \kv{} reuse framework for common serving scenarios. \method{} uses segment-level \kv{} reuse and management as its basic abstraction, introduces a \sparseq{} mechanism that is naturally aligned with the reuse structure, and uses sparse recomputation to combine context-independent segment-cache lookup with context-dependent runtime correction. Under complex, interleaved, and fragmented reuse patterns, \method{} maintains high reuse accuracy while adding little attention-computation complexity.

\begin{figure}[H]
    \centering
    \includegraphics[width=0.95\linewidth]{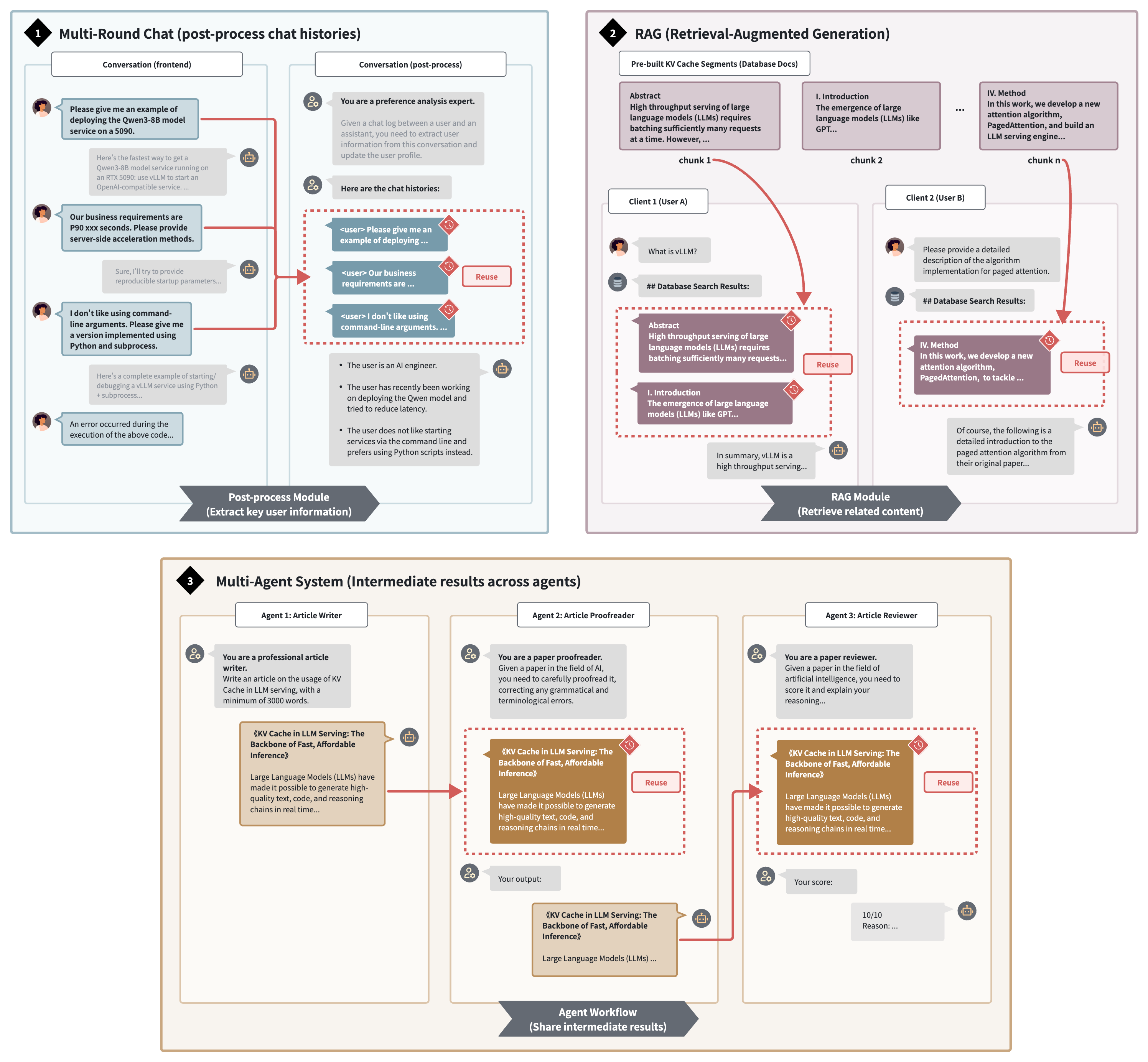}
    \caption{Three common \kv{} reuse scenarios supported by \method, including multi-round chat, RAG, and multi-agent systems. Reusable content appears as segment-level repeated text distributed across different positions in the prompt.}
    \label{fig:scenario}
\end{figure}

Figure~\ref{fig:scenario} shows three representative scenarios targeted by \method: historical dialogue reuse in multi-round chat, database document-segment reuse in RAG, and cross-agent intermediate-result reuse in multi-agent systems. These scenarios share a common property: repeated content does not necessarily appear as a prompt prefix, and instead appears as segments distributed across different positions.

\subsection{Limitations of Existing Methods}

Although several recent studies attempt to mitigate contextual coupling in arbitrary-position \kv{} reuse, we observe three key gaps that directly limit their applicability and engineering value in real systems.

\textbf{First, existing methods lack support for complex reuse patterns.}
Existing \kv{} reuse algorithms often implicitly or explicitly rely on specific input-structure assumptions, so that reuse decisions can be simplified into a small set of boundary conditions or concatenations of contiguous segments. For example, CacheBlend~\cite{yao2025cacheblend} targets RAG scenarios, concatenates multiple cached knowledge chunks into a new request, and selectively recomputes \kv{} tensors for a subset of tokens to restore cross-chunk interactions. This type of implementation is better suited to structures such as
\[
[\mathrm{Original}_1,\mathrm{Reuse}_1,\mathrm{Reuse}_2,\ldots,\mathrm{Reuse}_k,\mathrm{Original}_2].
\]
In real online traffic, however, a more common input form contains heavily interleaved reused and original segments:
\[
[\mathrm{Original}_1,\mathrm{Reuse}_1,\mathrm{Original}_2,\mathrm{Reuse}_2,\ldots,\mathrm{Original}_k].
\]
The length and semantic distribution of each original segment may be close to random, and the number of interleavings can grow substantially with multi-turn dialogue, RAG concatenation policies, and agent trajectories. For such patterns, methods that support only contiguous reused segments or depend on strict length and semantic-matching conditions can degrade to low hit rates across many requests, or can introduce poorly controlled representation error after a cache hit. More importantly, complex interleaving induces broader context-propagation effects during prefill. An original segment can affect the attention distribution and intermediate representation of multiple subsequent reused segments, making simple-similarity or local-patching strategies unstable. Therefore, the lack of systematic support for complex reuse structures is a primary reason why existing methods struggle to cover practical workloads.

\begin{figure}[H]
    \centering
    \includegraphics[width=0.85\linewidth]{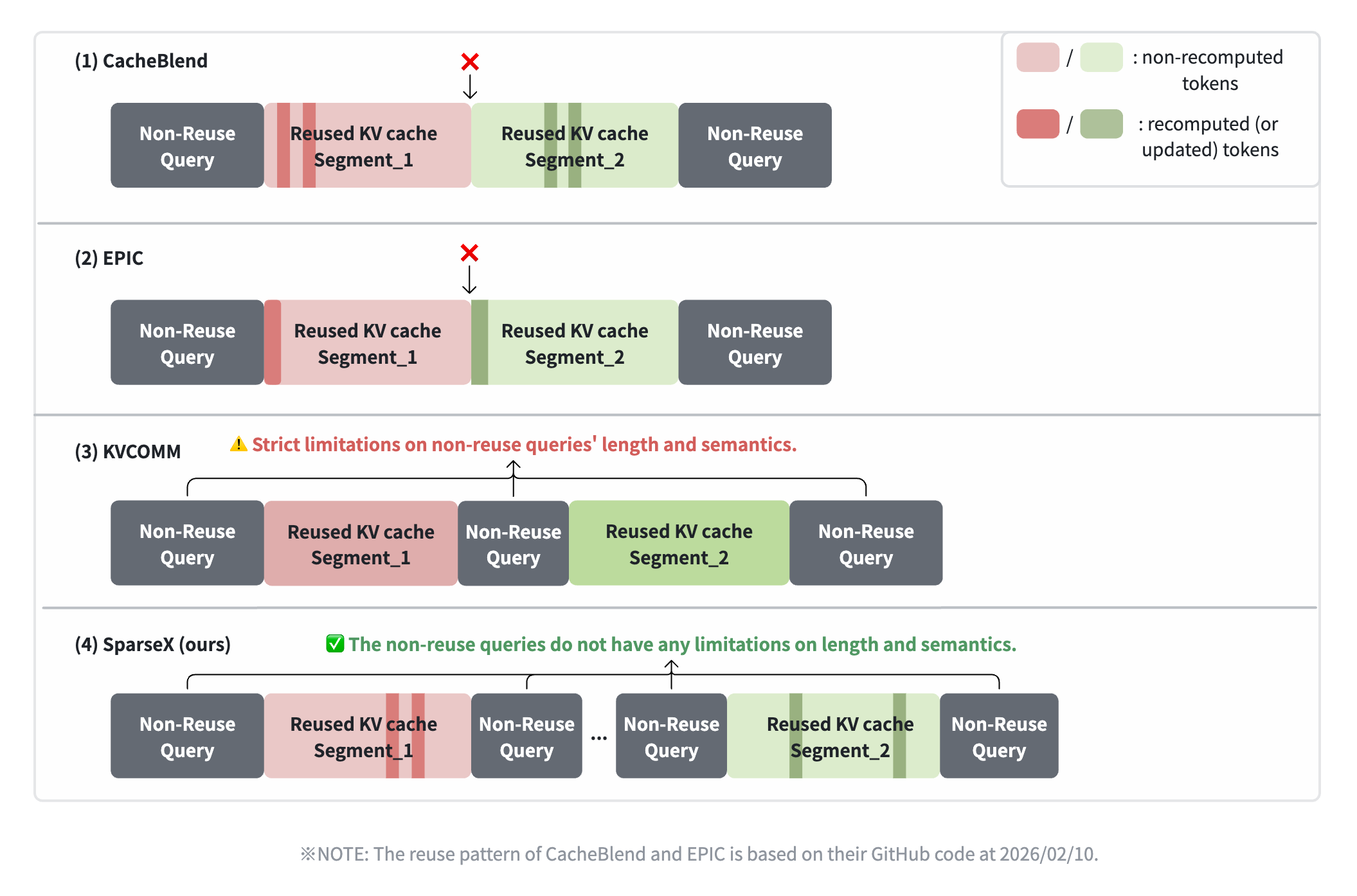}
    \caption{Comparison of different \kv{} reuse methods in terms of support for complex reuse patterns. \method{} supports arbitrary interleaving between reused segments and non-reuse query text.}
    \label{fig:reuse-pattern}
\end{figure}

Figure~\ref{fig:reuse-pattern} compares the applicability of different \kv{} reuse methods under complex interleaved inputs. CacheBlend and EPIC~\cite{hu2025epic} are better suited to more regular contiguous reuse structures. KVCOMM~\cite{ye2025kvcomm} imposes constraints on the length and semantics of non-reuse queries. \method{} allows non-reuse queries and reused segments to be arbitrarily interleaved in the prompt.

\textbf{Second, existing methods lack deep integration with inference engines.}
To obtain practical gains, \kv{} reuse must be co-designed with memory management, operator fusion, and scheduling paths in the inference engine. Otherwise, algorithmic savings can be offset by additional data movement, memory fragmentation, or incompatible execution backends. Some prior systems are only prototyped on older inference-engine versions. Other systems claim support for high-performance backends, but their key steps, such as selective token recomputation based on sparse attention, are often not incorporated into the execution graph of mainstream inference engines, which prevents them from sharing optimized paths with backends such as FlashAttention. Existing system implementations also commonly overlook compatibility with basic cache mechanisms such as Prefix Cache. If non-prefix reuse hits cannot cooperate with Prefix Cache, or if they cause a new request to exit the original cache framework, long-term hit rates and overall throughput can be reduced. More broadly, most \kv{} reuse methods have not entered the core ecosystem of mainstream inference engines such as vLLM because realistic serving constraints, including paged \kv{} management, batch scheduling, continuous request reuse, and GPU-memory watermark control, introduce non-negligible engine-side overhead that can shrink or destabilize the benefit.

\textbf{Third, existing methods target specific usage scenarios and lack generality.}
Prior work typically builds assumptions and evaluations around a single application paradigm, such as retrieval-fragment reuse in RAG or cross-context communication in agent systems. This scenario-specific route can show clear gains in paper evaluations, but it complicates maintenance and deployment in industrial systems. Practical services often contain multi-turn dialogue, RAG concatenation, tool calls, and multi-agent collaboration simultaneously, with input structures changing dynamically across requests and stages. If a reuse mechanism cannot share a unified loop for segment lookup, storage, reuse, and update across these scenarios, it is difficult to obtain stable gains in real workloads. We argue that an AI-infrastructure technique for \kv{} reuse should provide cross-scenario consistency: it should cover combinations of common serving patterns and preserve predictable quality and performance as the reuse pattern changes.

\subsection{Contributions}

To address these gaps, we propose \method: Efficient Segment-Level \kv{} Sharing in All Common Scenarios. The core observation behind \method{} is that, under \kv{} reuse and complex prompt concatenation, the model input naturally forms \sparseq{} indices that must perform full attention. These \sparseq{} indices can estimate which reused tokens are most important to the current request, thereby guiding subsequent selective recomputation.

We transform arbitrary-position reuse into a sparse-recomputation problem. \method{} organizes reuse objects at segment granularity, uses \sparseq{} to identify the token set that should be recomputed, and performs Sparse-KV Recomputation only for necessary tokens. This preserves high-accuracy reuse under complex interleaved reuse patterns. We also design and implement SparseX-vLLM, a system framework deeply compatible with vLLM. It uses PagedAttention for efficient \kv{} storage and reclamation, while remaining compatible with the FlashAttention backend and existing vLLM inference optimizations, including Prefix Cache.

Experimentally, we evaluate \method{} on three representative scenarios: multi-round chat, RAG, and agent workflows. The evaluation studies whether \method{} can significantly reduce prefill latency and improve end-to-end throughput while maintaining output quality close to full recomputation, and whether it outperforms existing methods under complex interleaved reuse patterns. Section~\ref{sec:evaluation} presents the full setup and results.

This paper makes the following contributions.

First, we propose \sparseq{}-driven token selection. We observe that, under segment-level \kv{} reuse, interleaving reused segments and original segments naturally creates a distinctive \sparseq{} pattern during the forward pass. Based on this observation, we propose a runtime sparse-pattern estimation algorithm. The algorithm follows the insight of MInference~\cite{jiang2024minference}, which uses last-Q indices to estimate sparse attention patterns, but replaces fixed last-Q probing with the \sparseq{} indices that naturally arise in \kv{} reuse workloads, yielding a more direct estimate of the tokens that require recomputation.

Second, we propose \method, a context-independent \kv{} reuse method based on \sparseq{} and Sparse-KV Recomputation. \method{} handles arbitrarily complex reuse patterns and improves reuse accuracy while adding little attention-computation complexity. The method is model-agnostic, training-free, and plug-and-play for decoder-only LLMs.

Third, we implement SparseX-vLLM, an efficient \kv{} reuse extension framework based on vLLM. SparseX-vLLM uses PagedAttention for efficient \kv{} management, remains compatible with FlashAttention backends, Prefix Cache, and the core inference optimizations in vLLM, and provides a unified segment lookup and storage mechanism for multi-round chat, RAG, and agent workflows, enabling rapid deployment of multiple models as online services.

\section{Related Work}

This section reviews three areas related to \method: \kv{} reuse, namely cross-request or cross-context \kv{} reuse and update; sparse attention, which provides sparse computation and \kv{}-budget management for long-context inference; and efficient attention backends and inference frameworks that support practical deployment, such as FlashAttention, FlashInfer~\cite{ye2025flashinfer}, vLLM, and SGLang~\cite{zheng2024sglang}.

\subsection{\kv{} Reuse}

\kv{} reuse aims to skip repeated prefill computation for token subsequences that have already appeared, directly reusing their Key/Value intermediate representations to reduce time-to-first-token and improve throughput. The most classical and widely deployed industrial form is Prefix Cache. When a new request shares an identical prefix token sequence with a historical request, the corresponding \kv{} tensors can be reused directly. Automatic Prefix Caching in vLLM implements prefix hits and reuse through block-level hashing and emphasizes that explicit tree structures are unnecessary for management. SGLang proposes RadixAttention, which maintains prompt and generated-token \kv{} tensors in a radix tree to support efficient prefix lookup, insertion, and eviction. Because the context is strictly identical, these methods can preserve output quality, but their hit condition strongly depends on token-level prefix equality and cannot cover more general arbitrary-position reuse.

Existing non-prefix \kv{} reuse methods can be roughly grouped into three categories.

The first category directly reuses or aligns \kv{} tensors at segment or chunk granularity without relying on sparse attention. EPIC proposes position-independent caching for multi-chunk inputs, typically in RAG settings. It reuses the \kv{} tensors independently computed for each chunk and introduces additional linking mechanisms to mitigate attention-sink issues caused by independent encoding. KVCOMM targets multi-agent workflows and proposes online cross-context KV-cache communication. When multiple agents share text under different prefix contexts, KVCOMM maintains an online anchor pool to estimate and correct \kv{} offset variance of shared content under different prefixes, enabling cross-agent \kv{} reuse. These methods usually incur low recomputation overhead, but they depend on stable prompt structures and predictable key-information locations, and they cannot fully restore cross-attention effects among segments. Section~\ref{sec:evaluation} further shows that inter-segment cross-attention is non-negligible on many tasks, and ignoring these interactions can cause visible degradation in generation quality.

The second category relies on auxiliary small models for token selection or importance prediction. CacheClip~\cite{yang2025cacheclip} observes that attention distributions in the last layer of a small model can approximate those of a large model, and uses this proxy to select tokens that are critical for restoring cross-chunk cross-attention. These tokens are then selectively recomputed. It also combines shared prefixes and grouping strategies to mitigate attention sinks and local-consistency issues. This route can improve the quality ceiling of token selection, but it introduces an additional model. In online serving, the extra model increases scheduling complexity: the system must switch between the large model and the small model, which can create pipeline bubbles and complicate batching when reuse and non-reuse requests are mixed under high concurrency.

The third category builds on local recomputation and sparse-attention ideas. CacheBlend targets multi-chunk RAG inputs and selectively recomputes a small subset of token \kv{} tensors to compensate for missing cross-chunk attention effects, while pipelining recomputation with \kv{} retrieval to reduce additional latency. DroidSpeak~\cite{liu2026droidspeak} studies the feasibility of sharing \kv{} tensors across different LLMs and, under the assumption of identical model architectures, selectively recomputes a small number of layers while reusing the remaining layers to support cross-model prefix \kv{} sharing. KVShare~\cite{yang2025kvshare} proposes Dual-Stage High Deviation for multi-tenant serving, conditionally selecting part of the \kv{} tensors for recomputation in both prefill and decode, and combining this with cache-aware scheduling for system efficiency. KVLink~\cite{yang2025kvlink} adjusts positional encoding and introduces trainable special tokens to mitigate attention bias after independently encoded documents are concatenated, representing a training-enhanced reuse direction. Among existing \kv{} reuse methods, sparse recomputation is the most promising route for balancing output quality and prefill efficiency.

\subsection{Sparse Attention}

Sparse attention studies how to approximate full attention under long contexts with sparse computation, reducing prefill or decode compute and memory cost. The works most related to \method{} use structural patterns in the attention matrix to construct sparse masks or sparse indices that can be implemented efficiently.

MInference analyzes and summarizes several common sparse-attention patterns for long-context prefill acceleration, including A-shape, Vertical-Slash, and Block-Sparse patterns. It proposes selecting a suitable pattern offline for each attention head and dynamically constructing sparse indices online according to the selected pattern, so customized GPU kernels can execute sparse attention efficiently. This work directly inspired our design of efficient sparse-index construction. The key difference is that MInference uses a fixed number of last-Q indices to estimate attention patterns, whereas \method{} uses \sparseq{} indices that naturally arise in \kv{} reuse workloads to estimate tokens requiring recomputation. These \sparseq{} indices often correspond to newly inserted instructions, queries, or routing text, and are therefore more directly related to how the current request uses reused segments.

A second route focuses on \kv{} budget management and compression. SnapKV~\cite{li2024snapkv} selects and clusters important \kv{} positions based on attention behavior, achieving training-free \kv{} compression and improving decode efficiency and memory efficiency under long sequences. H2O~\cite{zhang2023h2o} centers on heavy hitters and proposes a \kv{} eviction policy based on attention contribution, substantially reducing \kv{} memory footprint. StreamingLLM~\cite{xiao2024streamingllm} reveals the attention-sink phenomenon: retaining the \kv{} tensors of a small number of initial tokens can substantially recover the performance of sliding-window attention under very long sequences, supporting streaming long-context inference. These methods exploit sparsity in attention and context, but mainly optimize computation or storage for a single long-sequence request, instead of addressing arbitrary-position and arbitrary-pattern cross-request \kv{} reuse or restoring cross-segment interactions.

In particular, long-context Qwen models represented by Qwen2.5-1M~\cite{yang2025qwen25m} combine dual chunk attention with MInference in the model architecture. They use MInference sparse patterns to connect long text segments across chunks, achieving relatively stable length extrapolation. This direction shows that structured sparse attention can connect long-context fragments in practical LLMs, and it provides important motivation for introducing sparse attention into \kv{} reuse.

\subsection{Efficient Attention Backends and Inference Frameworks}

Efficient attention kernels and inference frameworks determine whether \kv{} reuse can be deployed with low engineering cost. The FlashAttention series implements high-throughput exact attention through IO-aware reordering and kernel fusion, becoming a foundational component for many inference systems and subsequent optimizations. FlashInfer provides a customizable attention engine and JIT-kernel system, and models multiple \kv{} layouts as block-sparse representations to support high-performance attention under cache organizations such as paged caches and radix-tree caches.

At the inference-framework level, vLLM proposes PagedAttention, which manages the \kv{} in pages to reduce memory fragmentation and support high-throughput continuous batching. Its Automatic Prefix Caching implements reuse of common system prompts or template prefixes through block-level hashes. SGLang proposes RadixAttention, which uses a radix tree to manage prompt and generated-token \kv{} tensors in a unified way, more systematically exploiting prefix sharing in structured execution. These systems and backends provide key support for SparseX-vLLM. \method{} must introduce generalizable segment-level reuse and sparse recomputation while preserving the efficiency of backends such as FlashAttention and FlashInfer and coexisting with Prefix Cache.

\section{Algorithm}

\subsection{RoPE Alignment of Segment-Level Cache}

We define Segment-Level Cache as a reuse abstraction that stores and shares the multi-layer Key/Value states produced during inference for contiguous token subsequences.

Formally, given a contiguous token segment
\[
S=(x_s,x_{s+1},\ldots,x_t),
\]
the \kv{} representation generated at layer \(\ell\) during an inference pass is
\[
\mathrm{KV}^{\ell}(S)\triangleq \bigl(K^{\ell}(S),V^{\ell}(S)\bigr),\qquad \ell\in[1,L],
\]
where
\[
K^{\ell}(S)\in\mathbb{R}^{|S|\times d_k},\qquad
V^{\ell}(S)\in\mathbb{R}^{|S|\times d_v}.
\]

Unlike traditional Prefix Cache, which only permits reuse of an identical prefix, Segment-Level Cache allows a previously computed segment \kv{} to be reused at arbitrary positions in a new sequence. This covers more general repeated structures in multi-round chat, RAG, and agent workflows. The key obstacle in cross-position reuse is positional-encoding mismatch: the absolute-position interval of a cached segment \(S\) in the old sequence usually differs from its absolute-position interval in the new sequence. For models using rotary position embedding (RoPE), Key vectors already carry absolute-position information when written into the \kv{}. Directly moving them to a new position causes an attention phase shift and breaks the consistency of subsequent computation.

\method{} removes this offset through RoPE alignment for reused segments, so cached segments can be reused at arbitrary positions. Suppose segment \(S\) occupies absolute positions \(\{p,p+1,\ldots,p+|S|-1\}\) in the original sequence and positions \(\{p',p'+1,\ldots,p'+|S|-1\}\) in the new sequence. For any layer \(\ell\), let \(k^\ell\) and \(q^\ell\) denote base vectors before RoPE, and let \(R_m\) denote the RoPE rotation matrix at position \(m\). The Key and Query written into or used with the \kv{} are
\[
\hat{k}_m^\ell=R_m k^\ell,\qquad
\hat{q}_m^\ell=R_m q^\ell.
\]

The key property of RoPE is that the attention score depends only on relative displacement. For the Query at position \(m\) and the Key at position \(n\),
\[
\begin{aligned}
\left\langle \hat{q}_m^\ell,\hat{k}_n^\ell \right\rangle
&=(R_m q^\ell)^\top(R_n k^\ell) \\
&=(q^\ell)^\top R_m^\top R_n k^\ell \\
&=(q^\ell)^\top R_{n-m} k^\ell.
\end{aligned}
\]
Therefore, when a cached Key is migrated from an old position \(n\) to a new position \(n'\), the Key corresponding to the new position can be obtained by applying an incremental rotation to \(\hat{k}_n^\ell\):
\[
\hat{k}_{n'}^\ell
=R_{n'}k^\ell
=R_{n'-n}R_n k^\ell
=R_{n'-n}\hat{k}_n^\ell.
\]
This transformation does not require explicitly recovering the unrotated \(k^\ell\); it can be performed directly in the cache domain. Standard RoPE acts only on Query and Key, so Value vectors carry no positional phase and can be copied directly.

Applying this derivation to the reuse of a contiguous segment \(S\), let the \(j\)-th token in the segment have old absolute position \(n=p+j\) and new absolute position \(n'=p'+j\). For all layers \(\ell\) and all \(j\in[0,|S|-1]\), \method{} performs
\[
\hat{k}_{p'+j}^{\ell}\leftarrow R_{p'-p}\hat{k}_{p+j}^{\ell},\qquad
\hat{v}_{p'+j}^{\ell}\leftarrow \hat{v}_{p+j}^{\ell}.
\]
Because all tokens within the same segment have the same displacement \(\Delta=p'-p\), segment-level alignment can be written in matrix form as
\[
K_{\mathrm{new}}^{\ell}(S)=\mathrm{Align}_{\Delta}\!\left(K_{\mathrm{old}}^{\ell}(S)\right),\qquad
V_{\mathrm{new}}^{\ell}(S)=V_{\mathrm{old}}^{\ell}(S).
\]
Here, \(\mathrm{Align}_{\Delta}\) applies the RoPE rotation \(R_\Delta\) to every \(d_k\)-dimensional Key vector in \(K_{\mathrm{old}}^{\ell}(S)\) according to the two-dimensional block structure of RoPE.

Computationally, the time complexity of this alignment process is
\[
O(|S|\cdot d_k),
\]
and its main cost is one read, rotation, and write of the cached Keys. To minimize engineering overhead, SparseX-vLLM implements a fused copy-alignment kernel that combines cached-page movement, Delta-RoPE rotation for Keys, and Value copying into a single GPU kernel. It completes RoPE alignment for all requests in a scheduling batch within a single vLLM engine core, maintaining stable throughput under high concurrency.

\subsection{\sparseq}

In segment-level \kv{} reuse, the input sequence can be represented as a token sequence \(x_{1:T}\) of length \(T\), consisting of multiple interleaved reuse segments and non-reuse segments. We use the index set
\[
\mathcal{I}_{\mathrm{reuse}}\subseteq\{1,\ldots,T\}
\]
to denote positions of all reused-segment tokens, and
\[
\mathcal{I}_{\mathrm{nr}}=\{1,\ldots,T\}\setminus\mathcal{I}_{\mathrm{reuse}}
\]
to denote all non-reuse token positions.

\begin{figure}[H]
    \centering
    \includegraphics[width=0.88\linewidth]{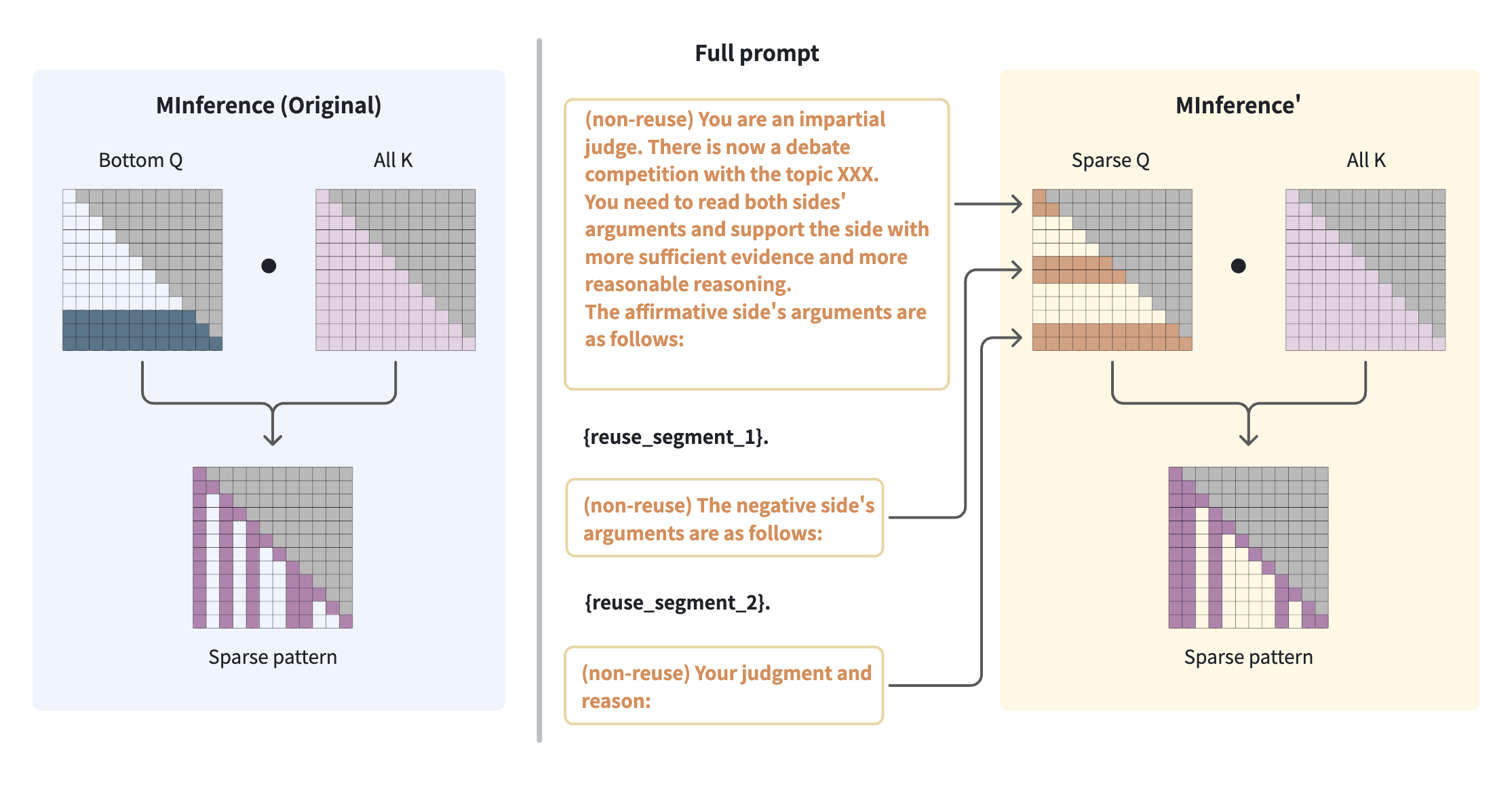}
    \caption{The \sparseq{} structure naturally formed in segment-level \kv{} reuse. Non-reuse segments must generate Query vectors and execute attention, while the \kv{} tensors of reuse segments can be read from the cache.}
    \label{fig:sparseq-structure}
\end{figure}

Figure~\ref{fig:sparseq-structure} shows the basic structure of \sparseq. The \kv{} tensors of reused segments have already been generated and stored by historical requests, so the current request only retrieves and position-aligns them. Non-reuse segments must execute forward computation again in the current context.

For a reused segment, the system already holds the \kv{} tensors of that segment at every layer, so full Query generation and attention computation for those tokens are unnecessary in the current prefill pass. By contrast, each non-reuse text segment, such as a new user input in multi-turn dialogue, a prompt instruction in RAG, or inserted instruction and routing text in an agent workflow, must fully compute its Query vectors and interact with all available Keys in the current context to produce correct hidden-state updates. This naturally creates a \sparseq{} structure during the forward pass: Query vectors that need to participate in attention exist only on \(\mathcal{I}_{\mathrm{nr}}\), while Query vectors on \(\mathcal{I}_{\mathrm{reuse}}\) can be omitted or selectively recomputed.

This structure enables \method{} to directly handle arbitrary interleavings of reuse segments and original segments. Because every non-reuse segment participates in causal attention at its true position, it can observe all previously aligned reused \kv{} tensors. Subsequent sparse recomputation then allows these non-reuse Query vectors to guide which reused tokens need correction. Thus, \sparseq{} both reduces unnecessary Query computation and provides a runtime signal for restoring cross-segment interactions under complex reuse patterns.

\method{} follows the idea in MInference of estimating sparse patterns from last-Q indices, but replaces the fixed tail Query set with the \sparseq{} indices naturally produced in reuse scenarios. Specifically, let
\[
Q_{\mathrm{sq}}\triangleq Q[\mathcal{I}_{\mathrm{nr}}]\in\mathbb{R}^{|\mathcal{I}_{\mathrm{nr}}|\times d}
\]
be the Query subset consisting only of non-reuse positions, and let \(K\in\mathbb{R}^{T\times d}\) be the Key matrix of the current sequence. We compute attention weights from \sparseq{} to all sequence Keys:
\begin{equation}
A_{\mathrm{sq}}
=\mathrm{softmax}\!\left(
\frac{Q_{\mathrm{sq}}K^\top}{\sqrt{d}}+
M_{\mathrm{causal}}[\mathcal{I}_{\mathrm{nr}},:]
\right),
\label{eq:sparseq-attn}
\end{equation}
where \(M_{\mathrm{causal}}\) is the causal mask and
\[
A_{\mathrm{sq}}\in\mathbb{R}^{|\mathcal{I}_{\mathrm{nr}}|\times T}.
\]
For each token position \(j\in\{1,\ldots,T\}\), we define the total attention intensity assigned by \sparseq{} as
\begin{equation}
s_j=\sum_{i=1}^{|\mathcal{I}_{\mathrm{nr}}|}
A_{\mathrm{sq}}[i,j].
\label{eq:sparseq-score}
\end{equation}
Finally, the top \(k\) positions are selected as
\begin{equation}
\mathcal{S}_{\mathrm{key}}
=\mathrm{TopK}\left(\{s_j\}_{j=1}^{T},k\right).
\label{eq:topk}
\end{equation}

With this procedure, \method{} uses the \sparseq{} region to estimate tokens that have important attention contribution in the input sequence. The complexity of Equation~\eqref{eq:sparseq-attn} is
\[
O(|\mathcal{I}_{\mathrm{nr}}|\cdot T\cdot d),
\]
the complexity of Equation~\eqref{eq:sparseq-score} is
\[
O(|\mathcal{I}_{\mathrm{nr}}|\cdot T),
\]
and top-\(k\) selection is
\[
O(T\log k)
\]
in common implementations. When the reuse ratio is high, \(|\mathcal{I}_{\mathrm{nr}}|\ll T\), so the estimation cost is much lower than the \(O(T^2d)\) cost of full attention. More importantly, \(Q_{\mathrm{sq}}\) corresponds to non-reuse Query vectors that must already be computed for the current request, so \method{} does not introduce an additional model or an independent dense probing stage for token selection.

At the implementation level, \method{} performs token selection using a global score aggregated across attention heads, instead of selecting patterns independently for each head as in the original MInference. This design aggregates \sparseq{} evidence across heads and produces a global token-importance estimate for the current request.

\begin{figure}[H]
    \centering
    \includegraphics[width=0.95\linewidth]{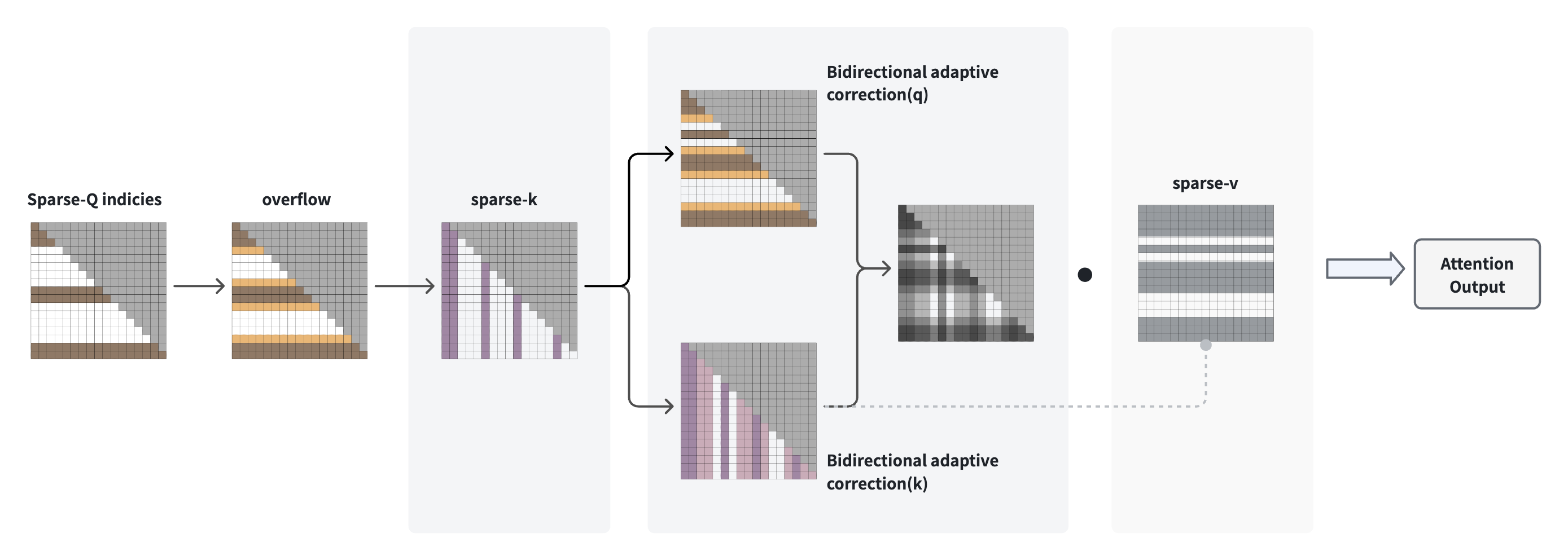}
    \caption{Key-token estimation based on \sparseq. \method{} replaces MInference's last-Q probing with \sparseq{} probing that naturally appears in \kv{} reuse workloads.}
    \label{fig:sparseq-selection}
\end{figure}

Figure~\ref{fig:sparseq-selection} illustrates \sparseq{}-driven token selection. MInference uses bottom-Q or last-Q vectors together with all Keys to estimate sparse patterns, whereas MInference' uses \sparseq{} vectors distributed at multiple prompt positions together with all Keys to estimate the Key tokens that require recomputation.

Unlike MInference, which estimates sparse attention patterns using a fixed number of last-Q tokens, non-reuse segments in \kv{} reuse workloads are often distributed among multiple reused segments. They specify how reused content should be interpreted, instruct the model on how to use reused text, and provide new reasoning goals. Our observation is that Query vectors from these non-reuse tokens more directly indicate which tokens in reused segments affect the current output. Therefore, \method{} uses \sparseq{} on \(\mathcal{I}_{\mathrm{nr}}\) to estimate the key tokens that should be corrected or recomputed first, yielding recomputation guidance suitable for arbitrary complex reuse structures.

As a fallback, when the prompt tail consists entirely of reused tokens and the tail region contains no \sparseq{} indices that must be computed, \method{} adds the last 64 query tokens of the final reuse segment to the recomputation set. In this case, \method{} falls back to a form close to last-Q probing in the original MInference, avoiding cases where reused tokens in some segments cannot obtain a recomputation pattern due to missing visible \sparseq{} tokens.

\subsection{Overflow}

The \sparseq{}-based token-selection method above selects globally important top-\(k\) tokens for the current request, forming a sparse structure similar to the vertical pattern in MInference. However, another common sparse-attention pattern, the slash pattern, is not always suitable for direct construction in \kv{} reuse. Slash patterns typically represent local relative-position attention and are built from similarity statistics between a set of Queries and all Keys. This implicitly requires obtaining the Key projection of every token first, which conflicts with the basic goal of segment-level \kv{} reuse: cached \(K,V\) tensors for reused segments should avoid full \(K/V\) projection and attention computation for those segments.

To resolve this conflict, \method{} adopts an overflow approximation for the local-neighborhood attention captured by the slash pattern. For each \sparseq{} index interval of a non-reuse segment, we expand outward by one block at both the beginning and the end of the interval. The last block of the previous reuse segment and the first block of the next reuse segment are included in the attention recomputation range.

The intuition is that the slash pattern mainly supplies local-neighborhood information in practice, while global key-token selection is handled by vertical patterns. The \sparseq{} region already performs dense computation within each non-reuse segment, covering the main dependency of its Query vectors on local context. Therefore, reused segments do not require an additional global periodic slash construction. Overflow further ensures that boundary information between non-reuse and adjacent reused segments is explicitly modeled, compensating for local attention missing most often at segment concatenation boundaries. This obtains benefits close to slash patterns without introducing full Key projection, while keeping implementation overhead low and stable.

\subsection{\fullsparse{} Hybrid Attention Based on a Layer-Specific Threshold}

Existing methods typically select top-\(k\) tokens for later sparse computation using attention results from the first layer or from a fixed number of early layers. For example, CacheBlend compares the first-layer \(K/V\) projection results with previously stored \kv{} tensors, selects tokens with large deviation, and gradually filters or recomputes these tokens in later layers. This mechanism assumes that early layers already provide token-importance signals that are reliable and insensitive to prompt structure.

\begin{figure}[H]
    \centering
    \includegraphics[width=0.95\linewidth]{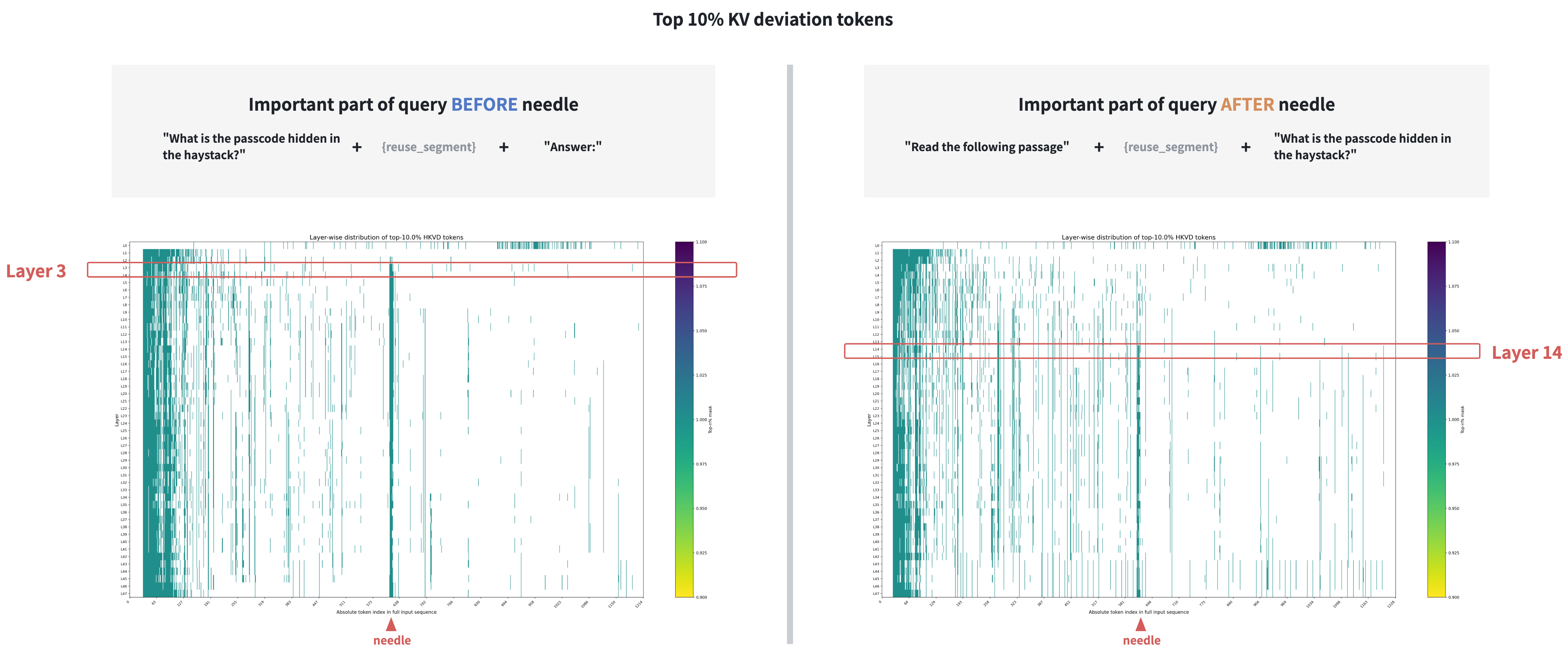}
    \caption{Top-token distribution based on \kv{} deviation in the NIAH task. The left panel places the important query cue before the needle, while the right panel places the important query cue after the needle.}
    \label{fig:kv-deviation}
\end{figure}

\begin{figure}[H]
    \centering
    \includegraphics[width=0.95\linewidth]{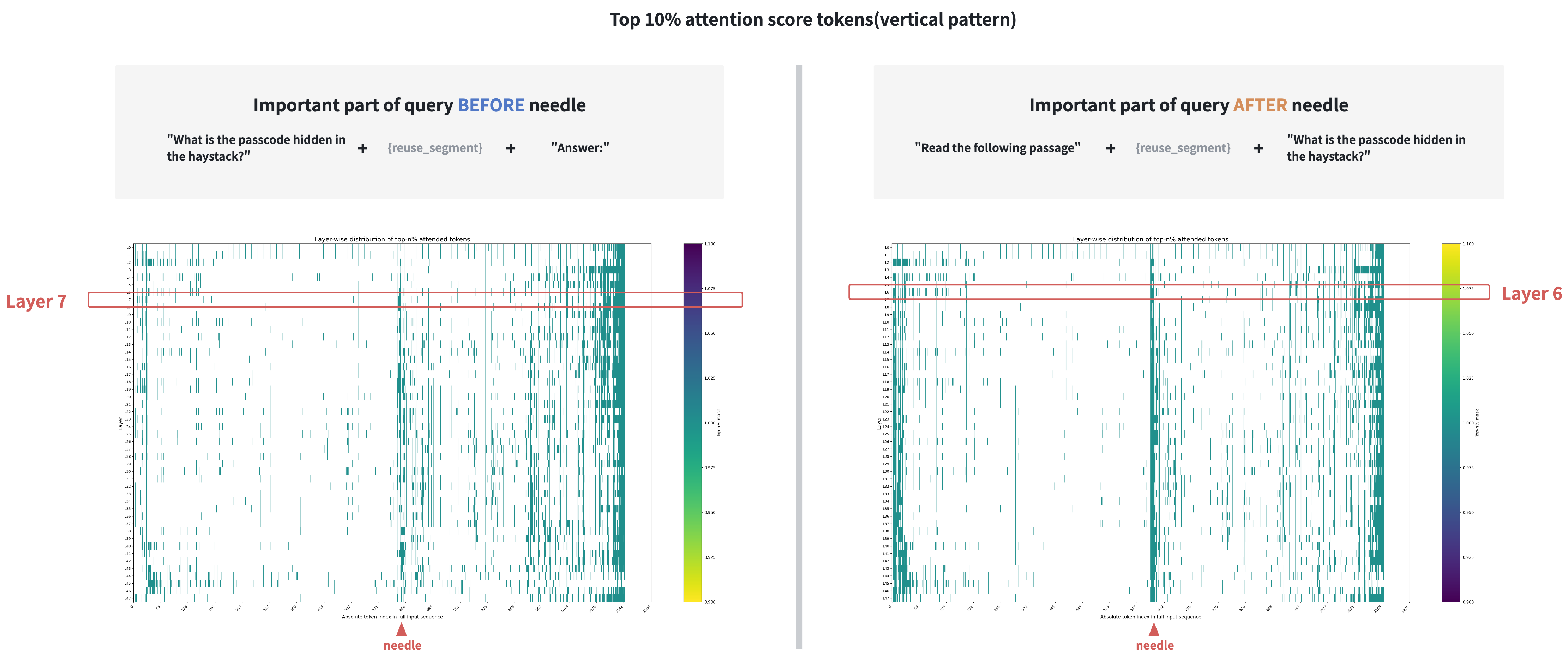}
    \caption{Top-token distribution based on last-query attention scores in the NIAH task. This estimation method forms a stable needle-localization signal early under both prompt layouts.}
    \label{fig:last-query}
\end{figure}

To examine this assumption, we conduct a layer-wise analysis on the needle-in-a-haystack task~\cite{kamradt2023niah}. Figures~\ref{fig:kv-deviation} and~\ref{fig:last-query} use Qwen3-30B-A3B as an example and compare two token-estimation signals. The first is based on \kv{} deviation, selecting tokens whose \kv{} representation changes substantially in the current context. The second is based on last-query attention scores, using the attention distribution from the tail query to all sequence Keys to identify important tokens in the vertical pattern. The two figures construct two prompt layouts: the important query cue appears before the needle, and the important query cue appears after the needle. The vertical axis denotes layer, the horizontal axis denotes absolute token position in the prompt, and the annotated needle region corresponds to the key information that should be stably identified.

Figure~\ref{fig:kv-deviation} shows that for CacheBlend-style \kv{} deviation methods and KVCOMM-style prompt-prefix embedding-similarity methods, the needle can be identified after a certain number of layers, but the stable layer position depends strongly on prompt construction. When the important query cue appears before the needle, the needle enters the top-deviation set at earlier layers. When the important query cue appears after the reused segment, the model requires more layers to form a stable needle signal. This indicates that \kv{} deviation reflects both the importance of the token itself and the context-propagation path induced by prompt layout. Under complex reuse patterns, relying on this signal at a fixed early layer for token selection cannot stably cover the key tokens that affect the output.

By contrast, Figure~\ref{fig:last-query} shows that last-query attention-score estimation is more stable with respect to prompt structure. Whether the important query cue appears before or after the needle, this method forms concentrated attention on the needle within a similar layer range. This observation is consistent with the MInference design of using last-Q indices to estimate vertical patterns: tail queries are closer to the final generation position and can aggregate information from preceding instructions, questions, and reused segments, so their attention distributions more directly indicate the Key tokens required by the current request. \method{} generalizes this idea to segment-level \kv{} reuse through \sparseq{}-driven estimation. When non-reuse queries are distributed across multiple prompt positions, \sparseq{} provides a selection signal more closely aligned with the current task goal than fixed early-layer \kv{} deviation.

Based on this observation, \method{} proposes a \fullsparse{} hybrid attention method with a layer-specific threshold to improve the stability of \sparseq{}-driven recomputation. We set a boundary in the early-to-middle layers. Before the boundary, the model uses full attention. At the boundary layer, the model performs one \sparseq{}-based key-token estimation, using \sparseq{} and the \kv{} tensors produced by earlier full-attention layers to estimate token indices that require sparse recomputation. After the boundary, all later layers use only \sparseq{} and the selected top-\(k\) tokens for Sparse-KV Recomputation. The remaining tokens directly reuse cached \kv{} tensors in these later layers.

Experiments across many tasks show that this layer boundary is not a fixed constant. Different model sizes, such as 7B and 72B, different downstream tasks, and different architectures within the same model family, such as dense Qwen3~\cite{yang2025qwen3} and Qwen3 MoE, require task-specific boundaries. For small dense models such as Qwen3-8B, the boundary is usually set to the first 15\% to 20\% of total layers. For larger models or MoE models, the boundary is usually close to the first 10\% to 15\% of total layers.

Combining Figures~\ref{fig:kv-deviation} and~\ref{fig:last-query}, model processing of key information in retrieval tasks can be roughly divided into three stages. The first stage corresponds to early layers, where the model mainly identifies structural information in input passages, such as segmentation, punctuation, and local formatting, without stable attention to a specific semantic range. The second stage corresponds to early-middle layers, where the model starts attending to the needle and roughly localizes token regions related to the question. At this stage, the attention map already reveals an approximate contour of the attention-concentrated region, although details are still adjusted in later layers. The third stage corresponds to late layers, where the model further processes semantic details and attention to the needle stabilizes. The \fullsparse{} hybrid attention of \method{} performs token selection near the second stage. At this point, the key region has an observable contour, and the overflow strategy provides moderate boundary slack for selected token intervals, covering local positional adjustments that may occur in later layers. This allows \method{} to avoid error propagation caused by overly early selection while preserving the benefit of reuse correction before all layers execute full attention.

Enabling \fullsparse{} hybrid attention introduces additional computation, so SparseX-vLLM makes it configurable. For simpler downstream tasks, the boundary from full attention to sparse attention can be set to a lower threshold, and sparse computation can even be enabled from layer 0. For more difficult long-context retrieval or complex reasoning tasks, hybrid attention usually improves quality substantially, especially on fine-grained retrieval subtasks in challenging benchmarks such as RULER, where token selection based only on the first layer is often insufficient for stable results.

\section{System Design}

\begin{figure}[H]
    \centering
    \includegraphics[width=0.9\linewidth]{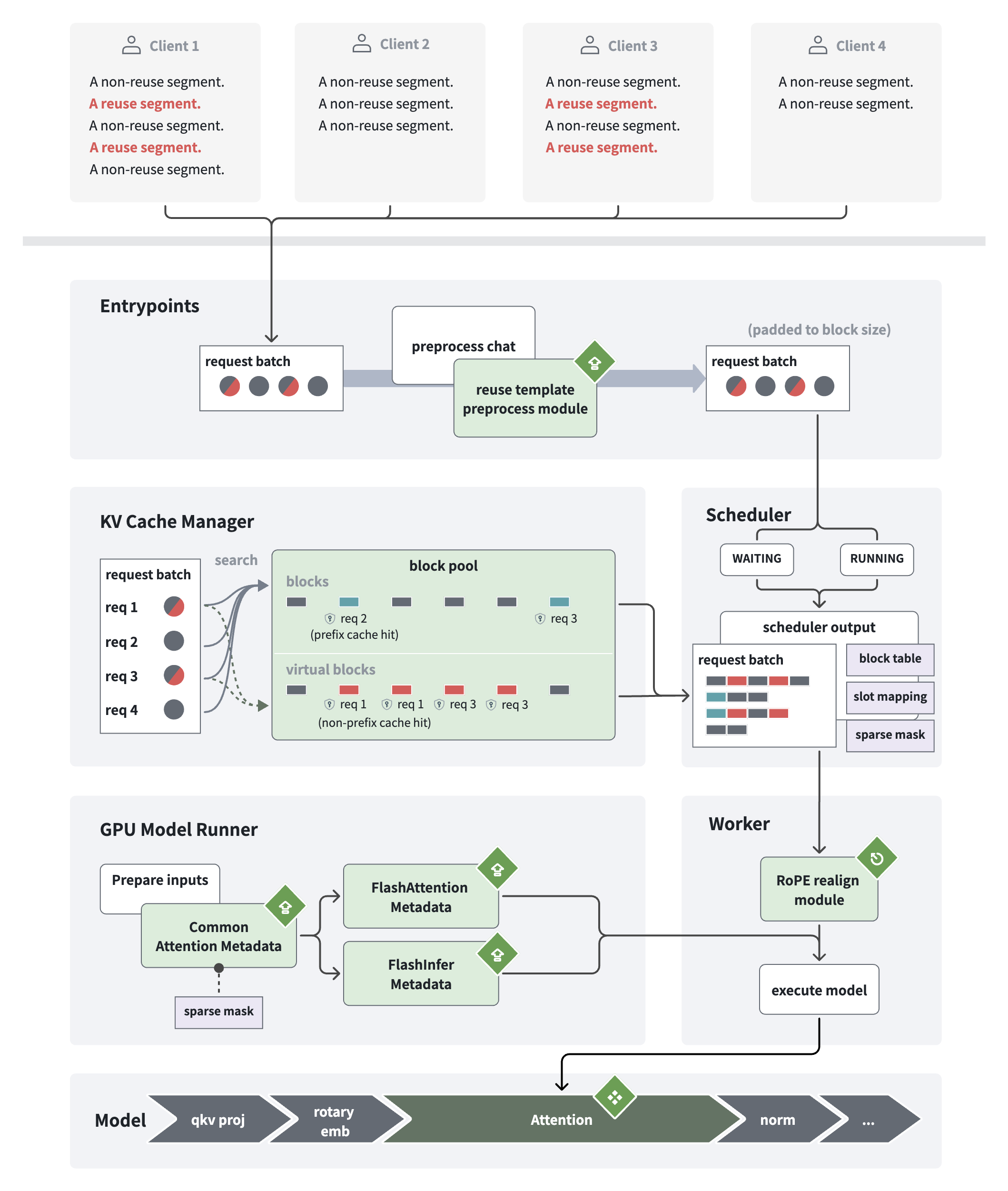}
    \caption{SparseX-vLLM system architecture. The system introduces \method{} modules into entrypoints, the \kv{} Cache Manager, Scheduler, Worker, GPU Model Runner, and model attention layers, while preserving vLLM's original PagedAttention and attention-backend execution path.}
    \label{fig:system}
\end{figure}

The goal of SparseX-vLLM is to integrate segment-level \kv{} sharing into the online inference path while preserving vLLM's original scheduling, PagedAttention management, and high-performance attention backends. Figure~\ref{fig:system} shows the overall architecture of SparseX-vLLM. A request first completes chat-template rendering, reuse-template preprocessing, and block-size padding at OpenAI-compatible entrypoints. The \kv{} Cache Manager then performs both ordinary Prefix Cache lookup and SparseX virtual-block lookup. The Scheduler writes hit results, block tables, slot mappings, and sparse masks into scheduler output. Before model execution, the Worker calls the RoPE realignment module to align reused \kv{} tensors. The GPU Model Runner constructs attention metadata based on the sparse mask and passes it to FlashAttention or FlashInfer backends. Finally, the model attention layers consume this metadata through an execution path compatible with original vLLM, unifying reuse and recomputation scheduling.

The SparseX-vLLM execution process discussed in this paper occurs entirely inside GPU memory within the vLLM engine. Its goal is to extend and optimize native vLLM Prefix Cache for non-prefix workloads. SparseX-vLLM is not a multi-level cache system that moves data across disk, CPU, and GPU, and it does not depend on swap or external storage movement for reuse. It is also compatible with such multi-level \kv{} management systems. In larger-scale deployments, \method{} can serve as a GPU-resident segment-reuse layer together with disk, CPU, or cross-node cache systems.

\subsection{Frozen Block}

\method{} introduces frozen blocks to support knowledge-base \kv{} management in RAG scenarios. A frozen block is a cached block obtained after knowledge-base text completes prefill inside vLLM and constructs \kv{} tensors. Its physical representation is identical to that of a native vLLM \kv{} block, with the same Key/Value tensor layout, block identifier, and reference-counting mechanism. Therefore, PagedAttention and the original block table can directly consume frozen blocks.

The main difference between frozen blocks and ordinary \kv{} blocks lies in lifecycle management. Ordinary blocks can be reclaimed by vLLM's native LRU or free-list policy after their reference count drops to zero. Blocks corresponding to knowledge-base text usually need to be reused across many requests over a long period. If they are fully managed by the original reclamation policy, they may be evicted too early during low-reference intervals, causing later RAG requests to miss the cache. \method{} marks such blocks as frozen blocks. When their reference count drops to zero, they do not automatically return to the free queue; instead, a dynamic resizing policy of the frozen-block pool selectively removes or re-adds them.

This design gives knowledge-base \kv{} tensors more stable residency while preserving vLLM block semantics. Frozen blocks change the caching, pinning, and release policy of blocks without changing the model execution path. To support online management, SparseX-vLLM exposes dedicated APIs for inserting and deleting entries in the frozen-block pool. Developers can add or remove blocks from the frozen-block pool in real time while the vLLM engine is running. In addition, when inference-engine load reaches a preset threshold, such as \kv{} block utilization exceeding 90\%, SparseX-vLLM can automatically delete the least-referenced blocks from the frozen-block pool to release GPU memory and avoid affecting ordinary online requests.

\subsection{Frozen Block Pool}

In RAG retrieval scenarios, SparseX-vLLM maintains a logical knowledge-base block pool inside the vLLM server, called the frozen-block pool. This pool does not correspond to an independently preallocated GPU memory region. It shares the same physical \kv{} blocks as the native vLLM block pool. The frozen-block pool only records which blocks belong to the knowledge base, which blocks can be reused by ordinary requests, and what fraction of the global block pool they occupy.

This design has two advantages. First, it avoids allocating a fixed-size GPU memory region solely for knowledge-base \kv{} tensors, reducing static memory-planning cost. The knowledge-base cache can grow or shrink in real time according to access frequency and reference state, without forcing the system to reserve a hard-to-adjust memory region for RAG at startup. Second, because frozen blocks and ordinary blocks share the same physical block pool, SparseX-vLLM can continue using vLLM's PagedAttention mechanism, block table, and slot mapping, without maintaining a separate attention memory layout for the knowledge-base cache.

To support both ordinary Prefix Cache and knowledge-base segment reuse, SparseX-vLLM uses different hash-matching strategies during lookup. The non-reuse prefix portion of a prompt continues to use native vLLM block hashes for Prefix Cache. Reuse segments are retrieved using hashes generated jointly from segment token IDs and extra keys. Thus, the same request can hit ordinary prefix blocks and frozen knowledge blocks within the same scheduling round, and these hits are subsequently organized into a unified block table consumable by the model.

\subsection{Extra Keys}

\method{} uses extra keys to distinguish cache namespaces and reduce false matches in segment-level block-hash lookup. Similar to native vLLM Prefix Cache, a block hash should not be determined only by token IDs, because identical token sequences can have different reuse semantics across business domains, knowledge bases, or user contexts. Therefore, when constructing a knowledge base, the system must specify an extra key for that knowledge base. A later request that wants to reuse a segment in that knowledge base must contain the same text and carry the same extra key.

The matching mechanism can be formalized as namespace-aware block lookup. Suppose a candidate reuse block has token sequence \(B=(x_i,\ldots,x_{i+b-1})\), where \(b\) is the block size, and extra key \(e\). \method{} computes its virtual hash as
\[
h_{\mathrm{virt}} = H(B,e),
\]
and looks up this hash in the corresponding virtual-block index or frozen-block pool. The block is considered reusable only when both token IDs and the extra key match. For ordinary prefix portions, the system continues to use the original vLLM prefix hash. For knowledge reuse or user-level reuse, the system uses a virtual hash with an extra key. In this way, \method{} maintains multiple logical cache namespaces within the same physical block pool, preventing RAG knowledge bases, user histories, and ordinary Prefix Cache from interfering with one another.

\subsection{Virtual Block}

Beyond knowledge-base blocks, many scenarios also need to reuse ordinary user chat histories or intermediate results from agents. \method{} does not introduce explicit user or session objects into vLLM, because doing so would change the abstraction boundary of a general inference engine. Instead, \method{} extends block-hash computation into user-level or business-level namespaces through extra keys and introduces virtual blocks to represent reusable cross-context segments.

A virtual block is a position-independent, prefix-independent, and context-independent representation of a block. Unlike a physical block in native vLLM, a virtual block does not occupy additional GPU \kv{} memory. Each virtual block corresponds to an existing physical block and maintains three pieces of information: a reference to the corresponding physical block, positional information of token IDs in that block, and a virtual hash generated from token IDs and extra keys. Since a virtual block only adds an index and pointer without copying Key/Value tensors, it does not increase GPU memory usage.

When a virtual block is selected for reuse, \method{} allocates an ordinary physical block for the current request as the carrier of \kv{} tensors used during inference, and copies the old \kv{} tensors corresponding to the virtual block into the new block. During copying, the Key tensor is RoPE-aligned according to the position difference between old and new contexts, while the Value tensor is copied directly. After this step, the new block has the same lifecycle and execution semantics as ordinary vLLM blocks and can be directly used by the subsequent block table, slot mapping, and attention backend.

This virtual-block design decouples index semantics from physical storage. The system can build multiple searchable views for the same physical block using different extra keys, while keeping a single copy of the \kv{} tensor in memory. When a request actually hits a reuse entry, the system creates a position-aligned physical copy for the current context.

\subsection{Block Retrieval and Reuse Mechanism}

The block retrieval and reuse mechanism of SparseX-vLLM is jointly implemented by entrypoints, the \kv{} Cache Manager, Scheduler, Worker, and GPU Model Runner.

When a request reaches the server, the client can carry an extra key for retrieval. Entrypoints first parse reuse segments in the prompt, complete chat-template preprocessing, and insert padding at necessary positions so that reuse-segment boundaries align with the vLLM block size. This step ensures that later segment lookup operates at block granularity and remains consistent with PagedAttention block management.

The \kv{} Cache Manager then performs two types of lookup for the request. For the non-reuse prefix portion, the system uses native vLLM Prefix Cache lookup to find exactly matched prefix \kv{} blocks. For reuse segments, the system retrieves corresponding virtual \kv{} blocks from the virtual-block list or frozen-block pool using segment token IDs and extra keys. If Prefix Cache misses, or if some tokens belong to non-reuse portions that must be recomputed in the current context, those tokens are separated and participate in prefill to build the current request's own \kv{} tensors.

After determining virtual-block hits, \method{} allocates the same number of ordinary blocks as carriers for \kv{} tensors used in the actual inference process. The Worker then invokes a fused copy-Delta-RoPE kernel to move \kv{} tensors from old blocks into newly allocated ordinary blocks. Key tensors are RoPE-corrected during copying, and Value tensors are copied directly. After this step, reused segments appear as ordinary vLLM \kv{} blocks in the current request, so they can be written into the block table and passed to the attention backend.

The Scheduler encodes ordinary Prefix Cache hits, SparseX virtual-block hits, and newly allocated blocks into a unified scheduler output. This output contains the request layout, block table, slot mapping, and sparse mask for the current batch. During input preparation, the GPU Model Runner reads this metadata and constructs both Common Attention Metadata and backend-specific metadata. Because the sparse mask and block table of \method{} are both integrated into vLLM's existing attention-metadata path, FlashAttention and FlashInfer backends can consume reused \kv{} tensors without changing interface semantics.

During the first prefill pass and subsequent decode, addresses of reused blocks are passed directly to the attention entry through the block table. For \sparseq{} tokens and selected key tokens that require recomputation, the model executes normal attention computation. For other reused tokens, the model directly reads the aligned \kv{} tensors. Thus, SparseX-vLLM realizes cooperative reuse among Prefix Cache, knowledge cache, and user-level virtual cache under a unified block pool, unified scheduler output, and unified attention-backend path.

\section{Evaluation}
\label{sec:evaluation}

This section evaluates \method{} in three scenarios: multi-round chat, long-context standard benchmarks, and multi-agent systems. All experiments run on an 8-GPU NVIDIA GeForce RTX 4090 server. The system implementation is based on vLLM v1 0.11.0, with SparseX-vLLM integrated into the scheduler, \kv{} Cache Manager, Worker, and attention-metadata path. For fair comparison, we compare only with open-source, reproducible \kv{} reuse methods that provide a vLLM implementation or can be reproduced on vLLM. For evaluation scenarios requiring LLM-as-a-judge, such as LOCOMO~\cite{maharana2024locomo}, LongMemEval~\cite{wu2025longmemeval}, and some RULER~\cite{hsieh2024ruler} subtasks, the judge model is GPT-5.1.

All experiments simulate realistic \kv{} reuse workloads. Specifically, we construct requests in two stages. The first stage sends a cache-building request, so reusable segments complete prefill inside the vLLM engine and are written into the block pool. The second stage sends an evaluation request, where these reusable segments are recombined with new non-reuse text according to the target scenario, and quality and TTFT are measured on that request. The full-recompute baseline denotes non-prefix workloads where the serving system cannot exploit Prefix Cache. Under such workloads, the default Prefix Cache in vLLM or SGLang usually cannot hit, so behavior is close to full prefill. This baseline therefore reveals the Prefix Cache blind spot targeted by \method{}.

\subsection{Multi-Round Chat}

We first evaluate \method{} in multi-round chat. This scenario is a canonical application of segment-level \kv{} reuse: historical dialogue content repeatedly appears across requests, while a new user input, system instruction, or generation prompt is usually inserted at different positions in the historical context, so the reuse structure no longer satisfies strict prefix sharing.

We use the LOCOMO and LongMemEval datasets. LOCOMO evaluates long-term conversational memory. Its samples usually contain long dialogue history, a user question, and information that must be extracted from multiple historical turns, testing whether a model can locate facts, preferences, events, and relations across long contexts. LongMemEval further emphasizes memory retrieval and multi-turn dialogue understanding over long time spans, requiring the model to recover fine-grained information relevant to the current question from a long context. Both datasets reflect the core challenge of history reuse in real multi-round chat systems: reused content is long, and the current request often depends on only a small number of scattered facts. Because the context length of LongMemEval exceeds the maximum context window of Qwen3-32B, we do not report LongMemEval results for Qwen3-32B.

Table~\ref{tab:chat} reports multi-round chat results. Full recompute provides the quality upper bound but has high TTFT. Naive reuse substantially reduces TTFT, but quality drops because it ignores contextual coupling between reused segments and the current query. CacheBlend and EPIC partially mitigate this issue, but they remain limited by their assumptions on reuse structure in complex multi-turn histories. By contrast, \method{} achieves scores closer to full recompute while retaining low TTFT. Enabling \fullsparse{} hybrid attention further improves \method{} on LOCOMO and LongMemEval, indicating that \sparseq{}-driven recomputation effectively handles session-history reuse in long-context multi-turn dialogue.

\begin{table}[H]
\centering
\caption{Quality and TTFT comparison on LOCOMO and LongMemEval in the multi-round chat scenario.}
\label{tab:chat}
\small
\begin{tabular}{llcccc}
\toprule
\multirow{2}{*}{\textbf{Model}} & \multirow{2}{*}{\textbf{Method}} & \multicolumn{2}{c}{\textbf{LOCOMO}} & \multicolumn{2}{c}{\textbf{LongMemEval}} \\
\cmidrule(lr){3-4}\cmidrule(lr){5-6}
 & & \textbf{Score} & \textbf{TTFT (s)} & \textbf{Score} & \textbf{TTFT (s)} \\
\midrule
\multirow{6}{*}{\textbf{Qwen3-32B}}
 & \textbf{full recompute} & 0.68 & 9.23 & -- & -- \\
 & \textbf{naive reuse} & 0.54 & 0.48 & -- & -- \\
 & \textbf{CacheBlend} & 0.60 & 1.32 & -- & -- \\
 & \textbf{EPIC} & 0.55 & 0.59 & -- & -- \\
 & \textbf{\makecell[l]{\method{} (w/o \fullsparse{}\\hybrid attention)}} & 0.64 & 0.66 & -- & -- \\
 & \textbf{\makecell[l]{\method{} (w/ \fullsparse{}\\hybrid attention)}} & 0.67 & 0.87 & -- & -- \\
\midrule
\multirow{6}{*}{\textbf{\makecell[l]{Qwen3-30B-\\A3B-Instruct-\\2507}}}
 & \textbf{full recompute} & 0.62 & 7.32 & 0.56 & 15.58 \\
 & \textbf{naive reuse} & 0.44 & 0.55 & 0.51 & 1.01 \\
 & \textbf{CacheBlend} & 0.59 & 0.73 & 0.50 & 2.48 \\
 & \textbf{EPIC} & 0.48 & 0.57 & 0.53 & 1.09 \\
 & \textbf{\makecell[l]{\method{} (w/o \fullsparse{}\\hybrid attention)}} & 0.57 & 0.56 & 0.56 & 1.15 \\
 & \textbf{\makecell[l]{\method{} (w/ \fullsparse{}\\hybrid attention)}} & 0.61 & 0.67 & 0.60 & 2.13 \\
\bottomrule
\end{tabular}
\end{table}

\subsection{RULER}

RULER is a standard benchmark for evaluating long-context capabilities. It contains multiple subtasks that measure whether models can perform retrieval, variable tracking, entity localization, and context integration in long prompts. Compared with a single question-answering dataset, RULER covers different types of long-range dependencies and systematically evaluates robustness under long-context and RAG-like input structures. We therefore use RULER to verify whether \method{} covers long-context retrieval and complex segment-reuse scenarios.

Table~\ref{tab:ruler} reports results on four RULER subtasks: MQ-NIAH, VT, CWE, and FWE. MQ-NIAH evaluates retrieval in multi-query needle-in-a-haystack settings. VT emphasizes variable tracking, requiring a model to track variable assignments and references in a long context. CWE and FWE evaluate extraction of entity- or fact-related information inside the context window. Together, these tasks test whether a \kv{} reuse method can preserve accuracy when key information appears at arbitrary prompt positions.

On the RULER variable-tracking task, \method{} with \fullsparse{} hybrid attention achieves accuracy close to full recompute, while other \kv{} reuse methods show lower accuracy. This result is consistent with the design of \method: \method{} does not assume that key information appears only at the beginning or end of a prompt, and instead uses \sparseq{} indices distributed across the prompt to estimate global key tokens. As a result, when key information crosses multiple segments and appears at arbitrary positions, \method{} can still localize tokens that require recomputation. Methods that rely on fixed boundaries, contiguous chunks, or early-layer signals are more likely to miss global key tokens when prompt structure deviates from their assumptions.

In Table~\ref{tab:ruler}, Average denotes the mean of the four RULER subtasks. Based on the table values, the average scores on Qwen3-8B for full recompute, naive reuse, CacheBlend, EPIC, \method{} without hybrid attention, and \method{} with hybrid attention are 0.96, 0.69, 0.81, 0.77, 0.88, and 0.94, respectively. On Qwen3-30B-A3B-Instruct-2507, the corresponding scores are 0.94, 0.70, 0.79, 0.76, 0.86, and 0.94. On Qwen3-32B, they are 0.97, 0.79, 0.86, 0.82, 0.91, and 0.98.

\begin{table}[H]
\centering
\caption{Accuracy comparison on RULER subtasks. Average is the mean of MQ-NIAH, VT, CWE, and FWE.}
\label{tab:ruler}
\small
\resizebox{\textwidth}{!}{
\begin{tabular}{llccccc}
\toprule
\textbf{Model} & \textbf{Method} & \textbf{MQ-NIAH} & \textbf{VT} & \textbf{CWE} & \textbf{FWE} & \textbf{Average} \\
\midrule
\multirow{6}{*}{\textbf{Qwen3-8B}}
 & \textbf{full recompute} & 1.00 & 1.00 & 0.95 & 0.87 & 0.96 \\
 & \textbf{naive reuse} & 0.69 & 0.33 & 0.92 & 0.82 & 0.69 \\
 & \textbf{CacheBlend} & 0.81 & 0.65 & 0.95 & 0.84 & 0.81 \\
 & \textbf{EPIC} & 0.84 & 0.49 & 0.93 & 0.83 & 0.77 \\
 & \textbf{\method{} (w/o hybrid attention)} & 0.90 & 0.80 & 0.95 & 0.86 & 0.88 \\
 & \textbf{\method{} (w/ hybrid attention)} & 1.00 & 0.94 & 0.94 & 0.89 & 0.94 \\
\midrule
\multirow{6}{*}{\textbf{\makecell[l]{Qwen3-30B-\\A3B-Instruct-\\2507}}}
 & \textbf{full recompute} & 1.00 & 0.98 & 0.93 & 0.86 & 0.94 \\
 & \textbf{naive reuse} & 0.73 & 0.37 & 0.90 & 0.79 & 0.70 \\
 & \textbf{CacheBlend} & 0.85 & 0.54 & 0.94 & 0.81 & 0.79 \\
 & \textbf{EPIC} & 0.84 & 0.45 & 0.96 & 0.79 & 0.76 \\
 & \textbf{\method{} (w/o hybrid attention)} & 0.86 & 0.79 & 0.96 & 0.83 & 0.86 \\
 & \textbf{\method{} (w/ hybrid attention)} & 1.00 & 0.97 & 0.95 & 0.83 & 0.94 \\
\midrule
\multirow{6}{*}{\textbf{Qwen3-32B}}
 & \textbf{full recompute} & 1.00 & 1.00 & 0.97 & 0.92 & 0.97 \\
 & \textbf{naive reuse} & 0.89 & 0.45 & 0.95 & 0.88 & 0.79 \\
 & \textbf{CacheBlend} & 0.93 & 0.66 & 0.96 & 0.89 & 0.86 \\
 & \textbf{EPIC} & 0.90 & 0.48 & 0.98 & 0.90 & 0.82 \\
 & \textbf{\method{} (w/o hybrid attention)} & 0.92 & 0.84 & 0.98 & 0.90 & 0.91 \\
 & \textbf{\method{} (w/ hybrid attention)} & 1.00 & 1.00 & 0.99 & 0.92 & 0.98 \\
\bottomrule
\end{tabular}
}
\end{table}

\subsection{Multi-Agent System}

We further evaluate \method{} on multiple agentic workflows. The downstream tasks cover a broad range of domains to verify applicability in collaborative reasoning systems, including MATH~\cite{hendrycks2021math}, GSM-Hard~\cite{gao2022pal}, and AQUA-RAT~\cite{ling2017aqua} for mathematical reasoning; SciBench~\cite{wang2024scibench} and GPQA~\cite{rein2024gpqa} for scientific reasoning; MMLU-Pro~\cite{wang2024mmluPro} for general knowledge understanding; and MedQA~\cite{jin2021medqa} and MedMCQA~\cite{pal2022medmcqa} for medical-domain evaluation. All evaluations are conducted in MASLab~\cite{ye2025maslab}, a unified benchmark platform for multi-agent systems.

To demonstrate practical value, we integrate \method{} into existing multi-agent frameworks, including AutoGen~\cite{wu2024autogen} and MAD~\cite{liang2024mad}, and evaluate applicability in realistic collaborative environments. Table~\ref{tab:agent} reports the full results. Overall, \method{} maintains quality close to vanilla execution across multiple models, tasks, and multi-agent workflows. On a few tasks, \method{} exhibits small performance drops, but these drops are usually limited. On more complex or knowledge-intensive tasks, such as GPQA, AQUA-RAT, and MedMCQA, \method{} also shows stable positive results. The larger Qwen2.5-72B Instruct model remains competitive after introducing \method, indicating that the method can improve efficiency while maintaining output quality. Consistent behavior across different agent frameworks further indicates that \method{} can serve as a system-level optimization for collaborative multi-agent inference.

\begin{table}[H]
\centering
\caption{Quality comparison across models, tasks, and workflows in the multi-agent system scenario.}
\label{tab:agent}
\scriptsize
\setlength{\tabcolsep}{3pt}
\renewcommand{\arraystretch}{0.94}
\resizebox{\textwidth}{!}{
\begin{tabular}{llcccccc}
\toprule
\multirow{2}{*}{\textbf{Model}} & \multirow{2}{*}{\textbf{Task}} & \multicolumn{3}{c}{\textbf{AutoGen}} & \multicolumn{3}{c}{\textbf{MAD}} \\
\cmidrule(lr){3-5}\cmidrule(lr){6-8}
 & & \textbf{Vanilla} & \textbf{\makecell{SparseX\\(w/o hybrid)}} & \textbf{\makecell{SparseX\\(w/ hybrid)}} & \textbf{Vanilla} & \textbf{\makecell{SparseX\\(w/o hybrid)}} & \textbf{\makecell{SparseX\\(w/ hybrid)}} \\
\midrule
\multirow{8}{*}{\textbf{\makecell[l]{Qwen2.5-7B\\Instruct}}}
 & \textbf{GSM-Hard} & 54.20 & 50.60 & 53.00 & 58.60 & 56.00 & 54.20 \\
 & \textbf{MATH} & 75.60 & 74.60 & 73.00 & 76.16 & 75.80 & 75.40 \\
 & \textbf{GPQA} & 30.36 & 28.12 & 29.02 & 25.00 & 30.13 & 25.67 \\
 & \textbf{AQUA-RAT} & 73.62 & 70.08 & 74.41 & 79.13 & 80.31 & 75.59 \\
 & \textbf{MMLU-Pro} & 55.20 & 53.00 & 52.80 & 42.20 & 46.20 & 50.00 \\
 & \textbf{SciBench} & 20.24 & 21.04 & 18.44 & 21.04 & 20.84 & 20.24 \\
 & \textbf{MedQA} & 57.40 & 57.60 & 57.60 & 33.20 & 46.40 & 53.20 \\
 & \textbf{MedMCQA} & 55.00 & 51.60 & 53.60 & 45.60 & 52.00 & 52.40 \\
\midrule
\multirow{8}{*}{\textbf{\makecell[l]{Qwen2.5-72B\\Instruct}}}
 & \textbf{GSM-Hard} & 65.00 & 65.80 & 64.40 & 62.40 & 63.60 & 62.40 \\
 & \textbf{MATH} & 82.00 & 79.80 & 79.20 & 83.00 & 83.80 & 82.40 \\
 & \textbf{GPQA} & 43.08 & 45.54 & 42.86 & 47.32 & 42.41 & 44.42 \\
 & \textbf{AQUA-RAT} & 77.56 & 80.31 & 78.35 & 79.53 & 79.53 & 79.92 \\
 & \textbf{MMLU-Pro} & 69.40 & 68.40 & 71.20 & 65.80 & 67.60 & 68.40 \\
 & \textbf{SciBench} & 25.05 & 26.05 & 26.05 & 27.25 & 26.85 & 27.45 \\
 & \textbf{MedQA} & 78.40 & 77.80 & 79.80 & 78.40 & 75.80 & 78.20 \\
 & \textbf{MedMCQA} & 70.20 & 68.00 & 68.40 & 66.40 & 67.60 & 64.20 \\
\midrule
\multirow{8}{*}{\textbf{\makecell[l]{Llama3-8B\\Instruct}}}
 & \textbf{GSM-Hard} & 14.60 & 12.00 & 25.60 & 22.60 & 29.40 & 28.00 \\
 & \textbf{MATH} & 16.80 & 15.60 & 35.80 & 33.80 & 38.40 & 37.20 \\
 & \textbf{GPQA} & 13.62 & 10.49 & 23.66 & 24.11 & 23.44 & 24.78 \\
 & \textbf{AQUA-RAT} & 48.82 & 51.18 & 50.00 & 46.85 & 58.66 & 59.45 \\
 & \textbf{MMLU-Pro} & 28.60 & 33.40 & 30.20 & 27.80 & 31.40 & 31.80 \\
 & \textbf{SciBench} & 6.01 & 4.21 & 8.62 & 11.02 & 11.22 & 11.22 \\
 & \textbf{MedQA} & 43.60 & 52.40 & 56.40 & 23.20 & 40.40 & 41.80 \\
 & \textbf{MedMCQA} & 41.60 & 47.20 & 45.80 & 25.80 & 36.40 & 34.80 \\
\bottomrule
\end{tabular}
}
\end{table}
\FloatBarrier

\section{Conclusion}

This paper presents \method, a segment-level \kv{} sharing method for common LLM serving scenarios. The core idea of \method{} is to use \sparseq{} indices that naturally arise in \kv{} reuse workloads to estimate key tokens requiring recomputation, and to combine this with RoPE alignment, overflow approximation, and \fullsparse{} hybrid attention to restore critical interactions between reused segments and the current context under complex interleaved reuse patterns. \method{} uses segments as the basic reuse unit, which allows a unified treatment of multi-round chat, RAG, and multi-agent workflows.

We further implement SparseX-vLLM on top of vLLM v1 0.11.0, integrating segment lookup, virtual blocks, frozen-block pools, position realignment, \sparseq{} selection, and attention-metadata construction into the original vLLM execution path. This implementation remains compatible with PagedAttention, Prefix Cache, and FlashAttention/FlashInfer backends, allowing \method{} to be deployed as a model-agnostic and training-free system extension for general LLM serving.

Experimental results show that \method{} maintains output quality close to full recompute while substantially reducing reuse overhead across multi-round chat, RULER long-context benchmarks, and multi-agent-system scenarios. In tasks where key information is distributed at arbitrary prompt positions, \sparseq{}-driven recomputation restores critical cross-segment interactions more stably than methods that rely on fixed boundaries or simple chunk concatenation. Future work can further study finer-grained adaptive recomputation ratios, backend support for more model architectures, and joint optimization with cross-node \kv{} storage and scheduling policies.

\clearpage
\appendix

\section{End-to-End SparseX Algorithm}
\label{app:sparsex-algorithm}

This appendix summarizes the end-to-end prefill procedure of \method{} in a framework-independent form. The algorithm assumes a sparse attention kernel that can natively consume an arbitrary recomputation set and can attend to a mixture of freshly recomputed \kv{} tensors and aligned cached \kv{} tensors. The purpose of the algorithm is to make explicit how segment lookup, RoPE alignment, \sparseq{}-driven token selection, overflow, \fullsparse{} hybrid attention, and first-token generation are connected in one prefill pass.

\begingroup
\small
\renewcommand{\arraystretch}{1.08}
\begin{longtable}{@{}p{0.045\linewidth}p{0.90\linewidth}@{}}
\caption{End-to-end \method{} prefill algorithm with segment-level \kv{} sharing.}
\label{tab:sparsex-end-to-end}\\
\toprule
\multicolumn{2}{@{}l}{\textbf{Algorithm 1} \method{} prefill with \sparseq{}-driven Sparse-KV Recomputation.} \\
\midrule
\endfirsthead
\toprule
\multicolumn{2}{@{}l}{\textbf{Algorithm 1} \method{} prefill with \sparseq{}-driven Sparse-KV Recomputation (continued).} \\
\midrule
\endhead
\midrule
\multicolumn{2}{r@{}}{\footnotesize Continued on next page} \\
\endfoot
\bottomrule
\endlastfoot
1: &
\textbf{Input:} prompt tokens \(x_{1:T}\); reusable segment candidates \(\mathcal{C}=\{S_m\}_{m=1}^{M}\); namespace keys \(e_m\); cached segment tensors \(\{\mathrm{KV}^{\ell}_{\mathrm{old}}(S_m)\}_{\ell=1}^{L}\); original segment positions \(p_m\); current segment positions \(p'_m\); layer boundary \(\ell^\star\); top-\(k\) budget \(k\); block size \(b\). \\
2: &
\textbf{Output:} first generated token \(y_1\) and updated \(\kv{}\) tensors for the current request. \\
3: &
\(\mathcal{I}_{\mathrm{reuse}}\leftarrow\emptyset\), \(\mathcal{I}_{\mathrm{nr}}\leftarrow\{1,\ldots,T\}\), \(\mathcal{B}_{\mathrm{hit}}\leftarrow\emptyset\). \\
4: &
\textbf{Segment lookup:} for each candidate segment \(S_m\), compute the virtual block key \(h_m=H(S_m,e_m)\); if the lookup succeeds, add its token indices to \(\mathcal{I}_{\mathrm{reuse}}\), remove them from \(\mathcal{I}_{\mathrm{nr}}\), and add the matched cached blocks to \(\mathcal{B}_{\mathrm{hit}}\). \\
5: &
Construct the \sparseq{} mask \(m_i\): \(m_i=1\) for \(i\in\mathcal{I}_{\mathrm{nr}}\) and \(m_i=0\) for \(i\in\mathcal{I}_{\mathrm{reuse}}\). Prefix-cache hits, if present, are excluded from the current prefill computation. \\
6: &
\textbf{RoPE alignment:} for every matched segment \(S_m\), let \(\Delta_m=p'_m-p_m\). For each layer \(\ell\in[1,L]\), write aligned tensors into the current request cache:
\[
K^{\ell}_{\mathrm{cur}}(S_m)\leftarrow \mathrm{Align}_{\Delta_m}\!\left(K^{\ell}_{\mathrm{old}}(S_m)\right),\qquad
V^{\ell}_{\mathrm{cur}}(S_m)\leftarrow V^{\ell}_{\mathrm{old}}(S_m).
\]
\\
\multicolumn{2}{@{}l}{\textbf{Phase 1. Dense contextualization before sparse selection.}} \\
7: &
\textbf{for} \(\ell=1,\ldots,\ell^\star\) \textbf{do} \\
8: &
\quad Execute full causal attention for the active prompt positions using the current mixture of newly computed tensors and aligned cached tensors. \\
9: &
\quad Update hidden states and write the newly produced \(K^{\ell}_{\mathrm{cur}},V^{\ell}_{\mathrm{cur}}\) for computed positions. \\
10: &
\textbf{end for} \\
\multicolumn{2}{@{}l}{\textbf{Phase 2. \sparseq{}-driven key-token estimation at the boundary.}} \\
11: &
Extract boundary-layer Query and Key tensors \(Q^{\ell^\star},K^{\ell^\star}\), and define
\[
Q_{\mathrm{sq}}\leftarrow Q^{\ell^\star}[\mathcal{I}_{\mathrm{nr}}].
\]
\\
12: &
Compute \sparseq{} attention to all visible Keys:
\[
A_{\mathrm{sq}}
=\mathrm{softmax}\!\left(
\frac{Q_{\mathrm{sq}}\left(K^{\ell^\star}\right)^{\top}}{\sqrt{d}}
+M_{\mathrm{causal}}[\mathcal{I}_{\mathrm{nr}},:]
\right).
\]
\\
13: &
Aggregate global key importance:
\[
s_j\leftarrow \sum_{i=1}^{|\mathcal{I}_{\mathrm{nr}}|}A_{\mathrm{sq}}[i,j],
\qquad j\in\{1,\ldots,T\}.
\]
\\
14: &
Select vertical key tokens:
\[
\mathcal{S}_{\mathrm{key}}\leftarrow \mathrm{TopK}\bigl(\{s_j\}_{j=1}^{T},k\bigr).
\]
\\
15: &
\textbf{Overflow:} for each contiguous interval of \(\mathcal{I}_{\mathrm{nr}}\), expand the interval by one neighboring block on both sides and collect the expanded positions as \(\mathcal{S}_{\mathrm{ov}}\). \\
16: &
\textbf{Tail fallback:} if the prompt tail contains only reused tokens and provides no visible \sparseq{} indices, add the last 64 query tokens of the final reused segment to \(\mathcal{S}_{\mathrm{tail}}\); otherwise set \(\mathcal{S}_{\mathrm{tail}}\leftarrow\emptyset\). \\
17: &
Define the final recomputation set for later layers:
\[
\mathcal{R}\leftarrow
\mathcal{I}_{\mathrm{nr}}
\cup \mathcal{S}_{\mathrm{key}}
\cup \mathcal{S}_{\mathrm{ov}}
\cup \mathcal{S}_{\mathrm{tail}}.
\]
\\
\multicolumn{2}{@{}l}{\textbf{Phase 3. Sparse-KV Recomputation after the boundary.}} \\
18: &
\textbf{for} \(\ell=\ell^\star+1,\ldots,L\) \textbf{do} \\
19: &
\quad Project \(Q^\ell_i,K^\ell_i,V^\ell_i\) only for tokens \(i\in\mathcal{R}\), and write the fresh \(K^\ell_i,V^\ell_i\) into the current request cache. \\
20: &
\quad For \(i\notin\mathcal{R}\), keep the RoPE-aligned cached \(K^\ell_i,V^\ell_i\) without recomputation. \\
21: &
\quad Run sparse causal attention with Queries from \(\mathcal{R}\) and Keys/Values from the full current context:
\[
O^\ell[\mathcal{R}]
\leftarrow
\mathrm{SparseAttn}\!\left(
Q^\ell[\mathcal{R}],
K^\ell_{\mathrm{cur}}[1:T],
V^\ell_{\mathrm{cur}}[1:T],
M_{\mathrm{causal}}
\right).
\]
\\
22: &
\quad Update hidden states only on \(\mathcal{R}\), then apply the layer output projection, residual path, normalization, and feed-forward module on the retained active states. \\
23: &
\textbf{end for} \\
\multicolumn{2}{@{}l}{\textbf{Phase 4. First-token generation.}} \\
24: &
Read the final hidden state associated with the last active prompt position \(t_{\mathrm{out}}\), compute logits \(z=W_{\mathrm{lm}}h^L_{t_{\mathrm{out}}}\), and sample or decode \(y_1\sim\mathrm{Decode}(z)\). \\
\end{longtable}
\noindent{\footnotesize \textit{Note.} The table uses \(T\) for prompt length, \(L\) for the number of layers, \(d\) for attention head dimension, \(b\) for block size, and \(\ell^\star\) for the configurable transition layer of \fullsparse{} hybrid attention. The \sparseq{} estimation step has complexity \(O(|\mathcal{I}_{\mathrm{nr}}|Td)\) for attention-score computation and \(O(T\log k)\) for top-\(k\) selection under standard implementations. RoPE alignment for a reused segment \(S\) costs \(O(|S|d_k)\) per layer. After the boundary, attention is evaluated only for \(|\mathcal{R}|\) recomputed query positions while reading Keys and Values from the full current context.}
\par
\endgroup

Algorithm~\ref{tab:sparsex-end-to-end} presents \method{} as a single prefill pipeline. Segment lookup first determines which blocks can be reused under namespace-aware keys, and RoPE alignment converts the cached Keys to their current absolute positions. The early full-attention phase provides a stable contextual signal before token selection. At the boundary layer, \sparseq{} attention estimates globally important Key tokens, overflow adds local boundary coverage around non-reuse intervals, and the last-64 fallback handles prompts whose tail contains only reused content. Later layers recompute only the union of non-reuse tokens, selected vertical tokens, overflow tokens, and fallback tail tokens, while all remaining reused tokens keep their aligned cached \kv{} tensors. This procedure preserves the end-to-end causal prefill interface while reducing redundant computation over reused segments.

\clearpage
\section{Evaluation Prompt Construction Details}
\label{app:evaluation-prompts}

This appendix describes how the prompts used in our evaluation are constructed for the \kv{} reuse setting. Each example follows the same two-phase structure. In Phase 1, the complete prompt or reusable segment is executed once to build reusable \kv{} blocks. In Phase 2, the new request is decomposed into a prompt prefix, one or more reusable prompt-body segments, and a prompt suffix. The reusable prompt-body regions are loaded from previously built \kv{} blocks, while the prefix, suffix, and required boundary regions provide the \sparseq{} indices used by \method{}.

Across the following figures, the pink regions denote original tokens that are not directly reused and therefore induce \sparseq{} positions. The reusable prompt body is copied from cached blocks after RoPE alignment. When a cached block is incomplete under vLLM's block lifecycle, the incomplete tail block is recomputed and also contributes natural \sparseq{} positions.

\subsection{LongMemEval and LOCOMO}

For LongMemEval and LOCOMO, Phase 1 builds \kv{} blocks from a complete conversation-history sample. The context length is one LongMemEval or LOCOMO sample. The prompt contains many irrelevant chat histories and a smaller number of key chat histories. In Phase 2, the request uses a fixed instruction prefix asking the model to read conversation histories and answer based on the histories and user memories. The prompt body is the reused conversation history, and the suffix contains the question. Because vLLM does not cache the final incomplete assistant block before the end-of-sequence token, that block is recomputed in the reuse phase and becomes part of the natural \sparseq{} region.

\begin{figure}[H]
    \centering
    \includegraphics[width=\linewidth]{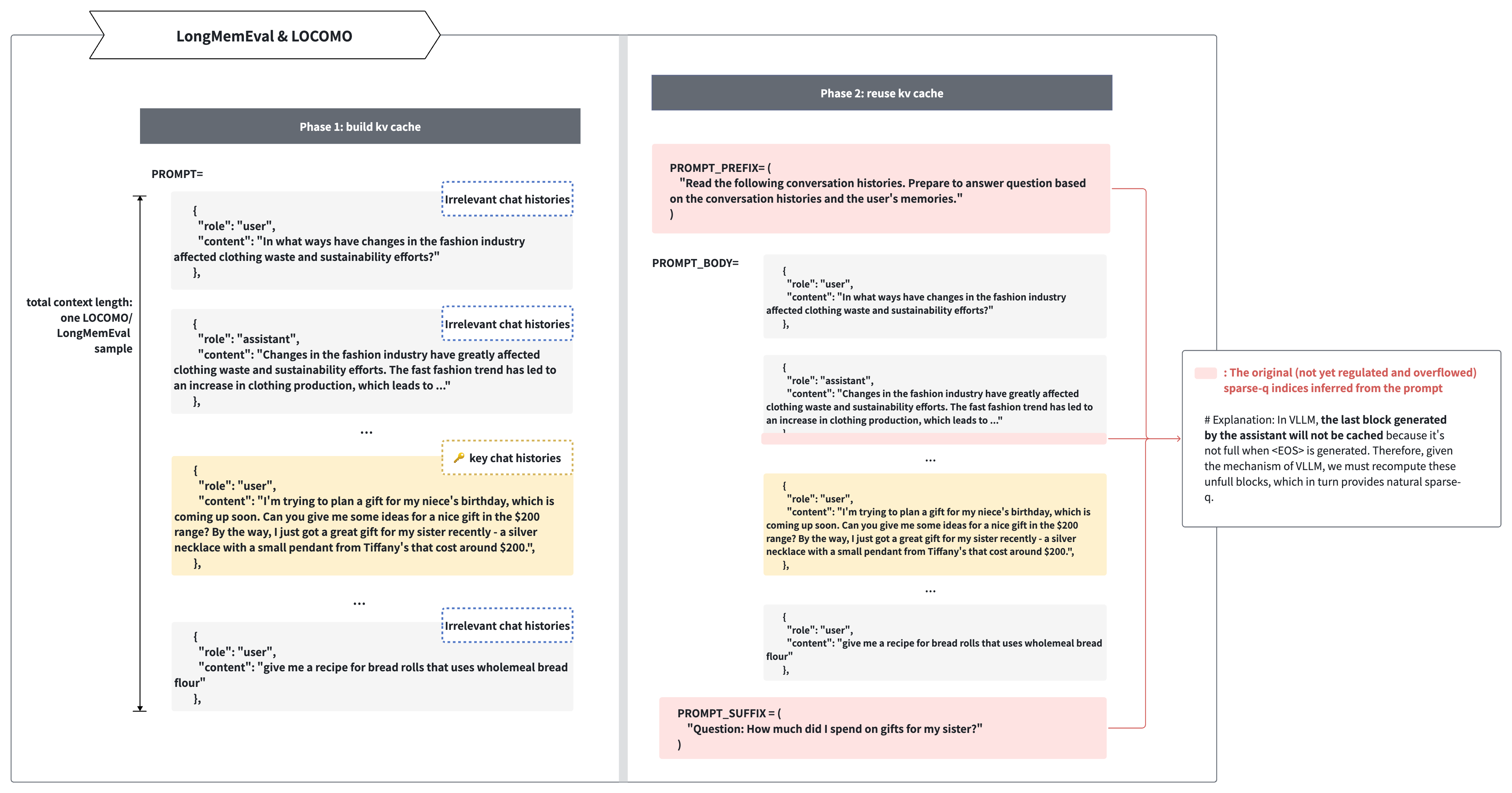}
    \caption{Prompt construction for LongMemEval and LOCOMO.}
    \label{fig:app-locomo-lme}
\end{figure}

\subsection{MQ-NIAH}

MQ-NIAH uses a 16K-token context composed of multiple 4K-token segments. Each segment contains noise paragraphs and one hidden needle. Phase 1 builds reusable \kv{} blocks for the complete prompt containing all needle-bearing segments. In Phase 2, the prefix states that special numbers are hidden in the text and asks for all special numbers associated with the queried entities. The prompt body is reused from cached noise-and-needle segments. The suffix asks the model to answer with the special numbers mentioned in the provided text.

\begin{figure}[H]
    \centering
    \includegraphics[width=\linewidth]{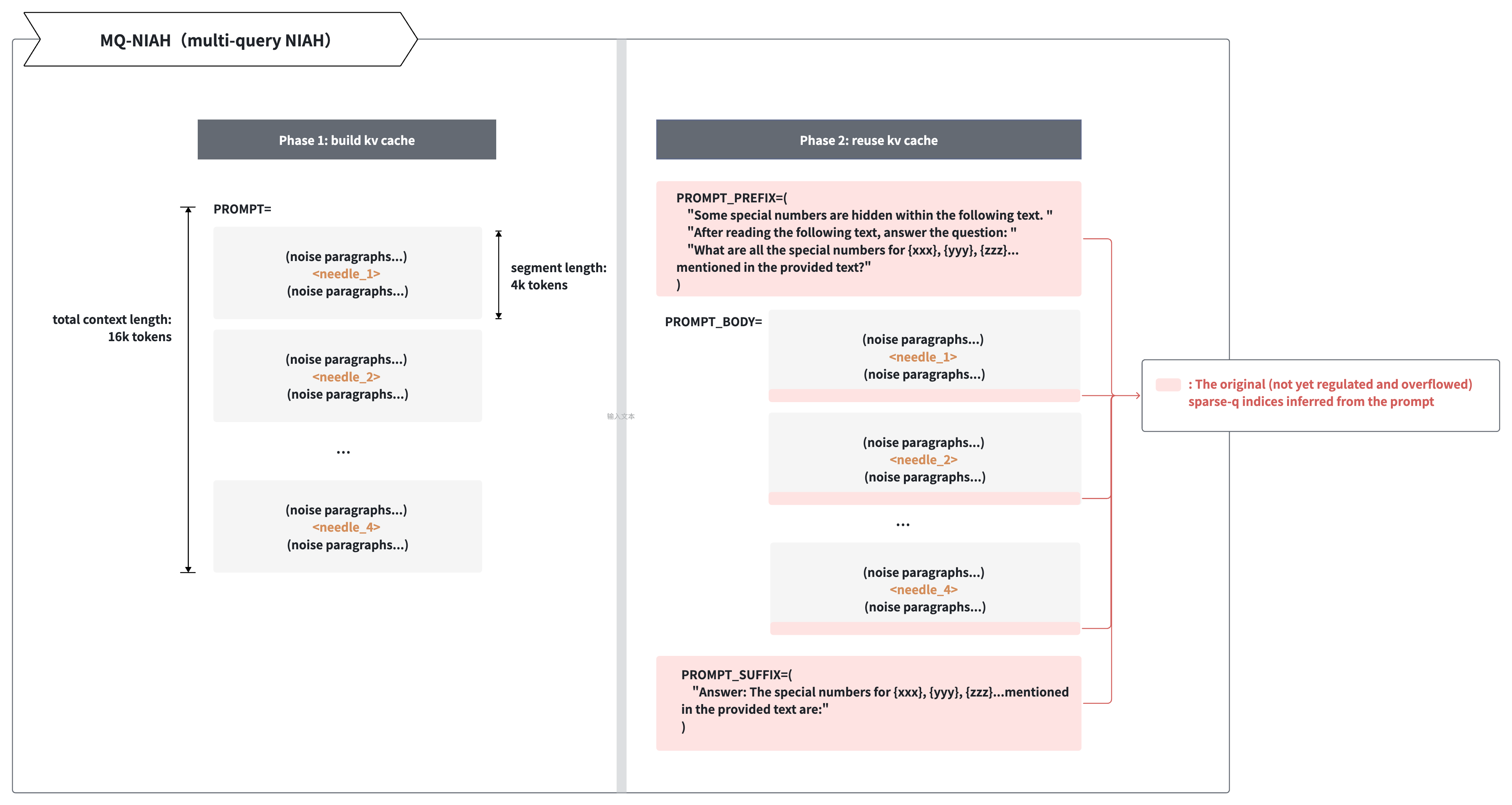}
    \caption{Prompt construction for MQ-NIAH.}
    \label{fig:app-mq-niah}
\end{figure}

\subsection{Variable Tracking}

The variable-tracking task also uses a 16K-token context split into 4K-token segments. Each segment contains noise paragraphs and variable-assignment statements such as assignments to a base value or assignments through previous variables. Phase 1 builds \kv{} blocks over the full variable-chain prompt. In Phase 2, the prefix instructs the model to memorize and track variable-assignment chains and to output all variables assigned to a target base value. The prompt body reuses the variable-assignment segments, and the suffix asks for the final list of variables.

\begin{figure}[H]
    \centering
    \includegraphics[width=\linewidth]{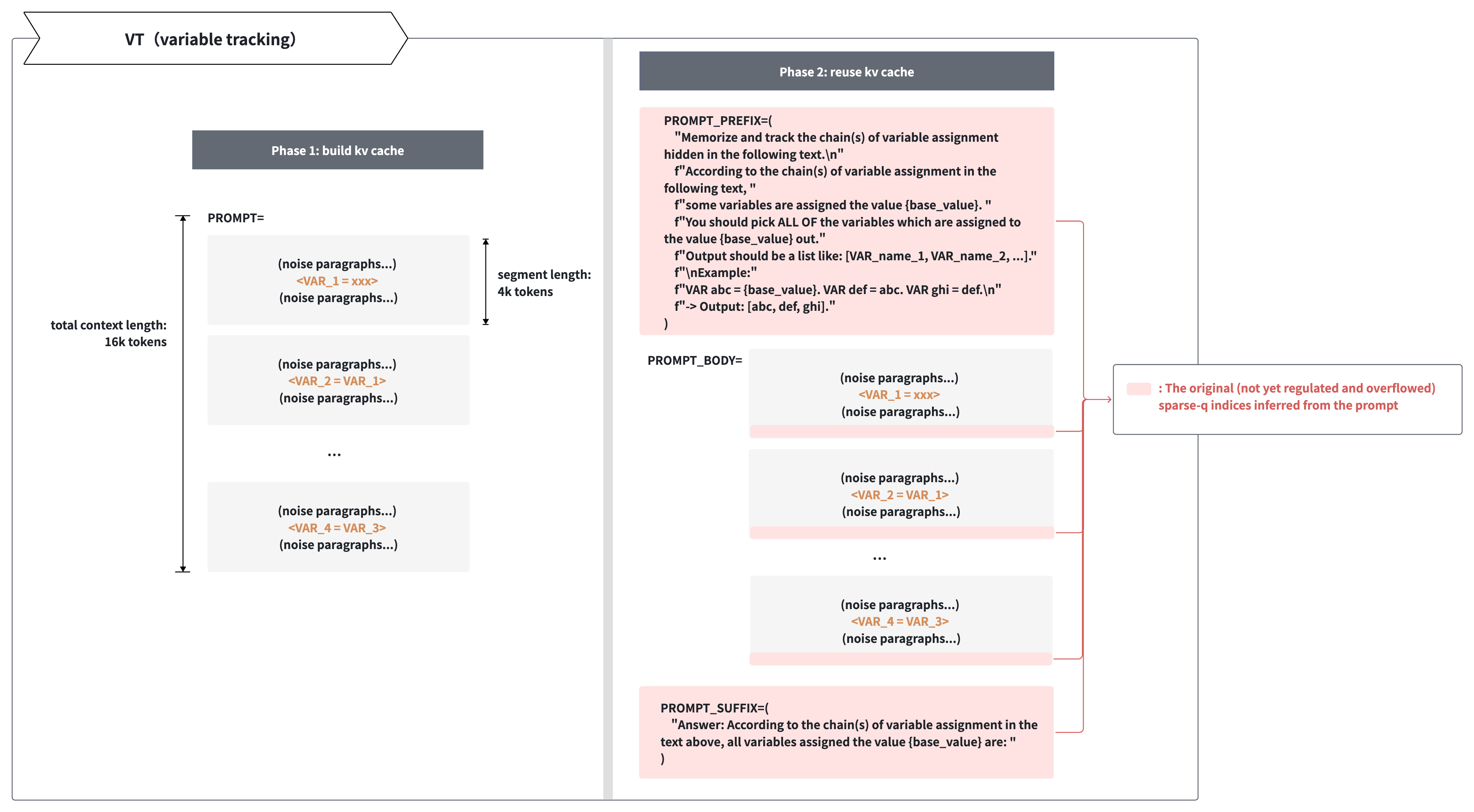}
    \caption{Prompt construction for variable tracking.}
    \label{fig:app-vt}
\end{figure}

\subsection{Common Words Extraction}

For common-words extraction, the total context length is 16K tokens, and each reused segment is approximately 4K tokens. The prompt body contains repeated numbered word lists mixed with random words. Phase 1 builds \kv{} blocks from the full word-list prompt. In Phase 2, the prefix provides an example and asks the model to identify the ten most common words in the list. The prompt body is loaded from cached word-list segments, and the suffix asks the final question.

\begin{figure}[H]
    \centering
    \includegraphics[width=\linewidth]{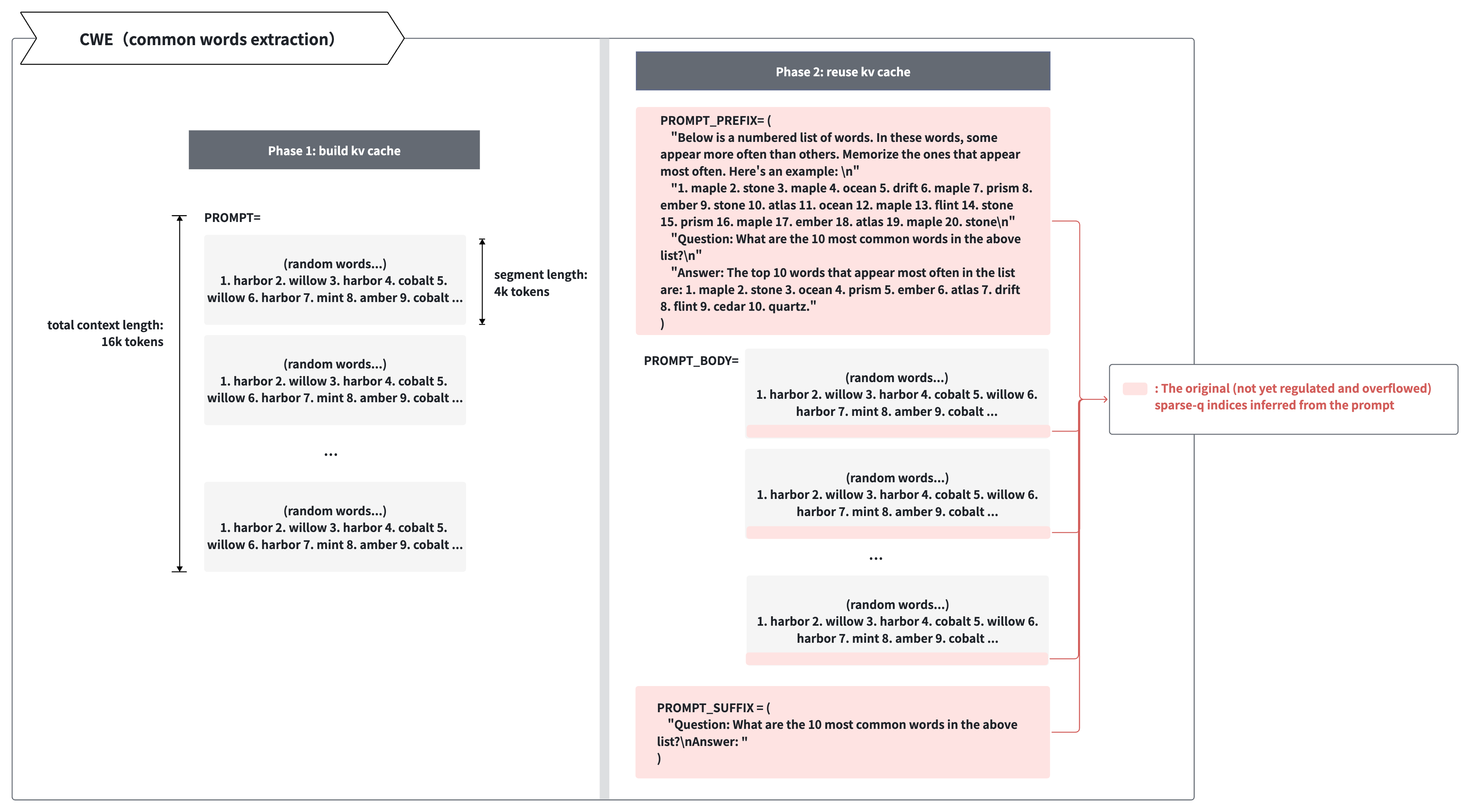}
    \caption{Prompt construction for common words extraction.}
    \label{fig:app-cwe}
\end{figure}

\subsection{Frequent Words Extraction}

For frequent-words extraction, the total context length is 16K tokens with 4K-token reusable segments. The prompt body contains coded words in random text, and the task is to track word frequency. Phase 1 builds reusable \kv{} blocks over the full coded-text prompt. In Phase 2, the prefix asks the model to read the coded text and find the top-\(k\) most frequently appearing coded words. The prompt body is reused from the cached coded-text segments, and the suffix requests the final answer without additional explanation.

\begin{figure}[H]
    \centering
    \includegraphics[width=\linewidth]{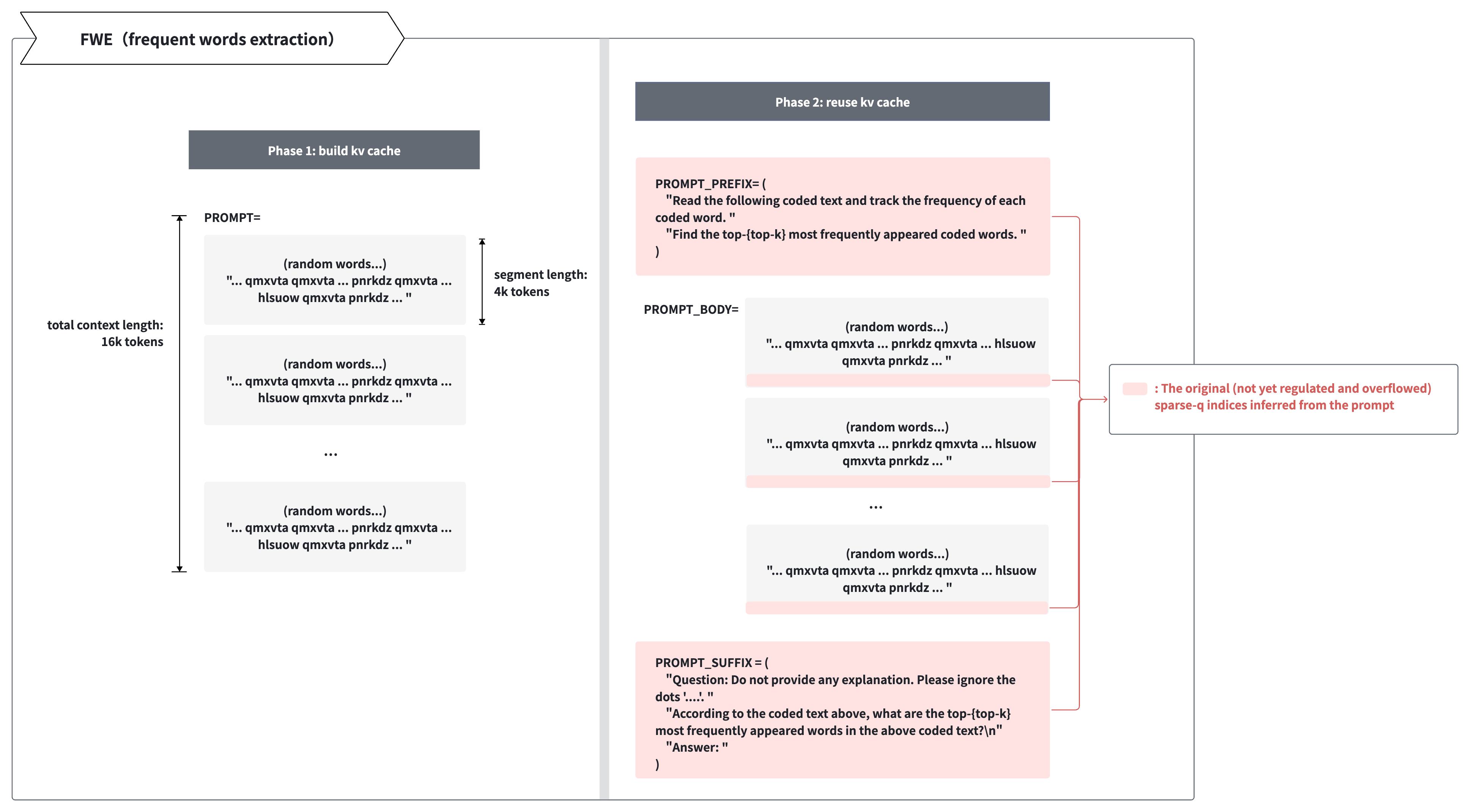}
    \caption{Prompt construction for frequent words extraction.}
    \label{fig:app-fwe}
\end{figure}

\subsection{AutoGen}

For AutoGen-style agent workflows, the total context length is determined by LLM generation and is usually no more than 6K tokens. Phase 1 builds reusable \kv{} blocks from agent messages, including the system instruction, the user problem, and generated outputs from previous agents. In Phase 2, the prefix asks a mathematical analyst to evaluate responses from different agents. The prompt body inserts the reused outputs of Agent 1 through Agent \(n\), and the suffix asks for the final evaluation and answer. The generated agent-output blocks are the reused segments, while the surrounding task instructions provide the \sparseq{} regions.

\begin{figure}[H]
    \centering
    \includegraphics[width=\linewidth]{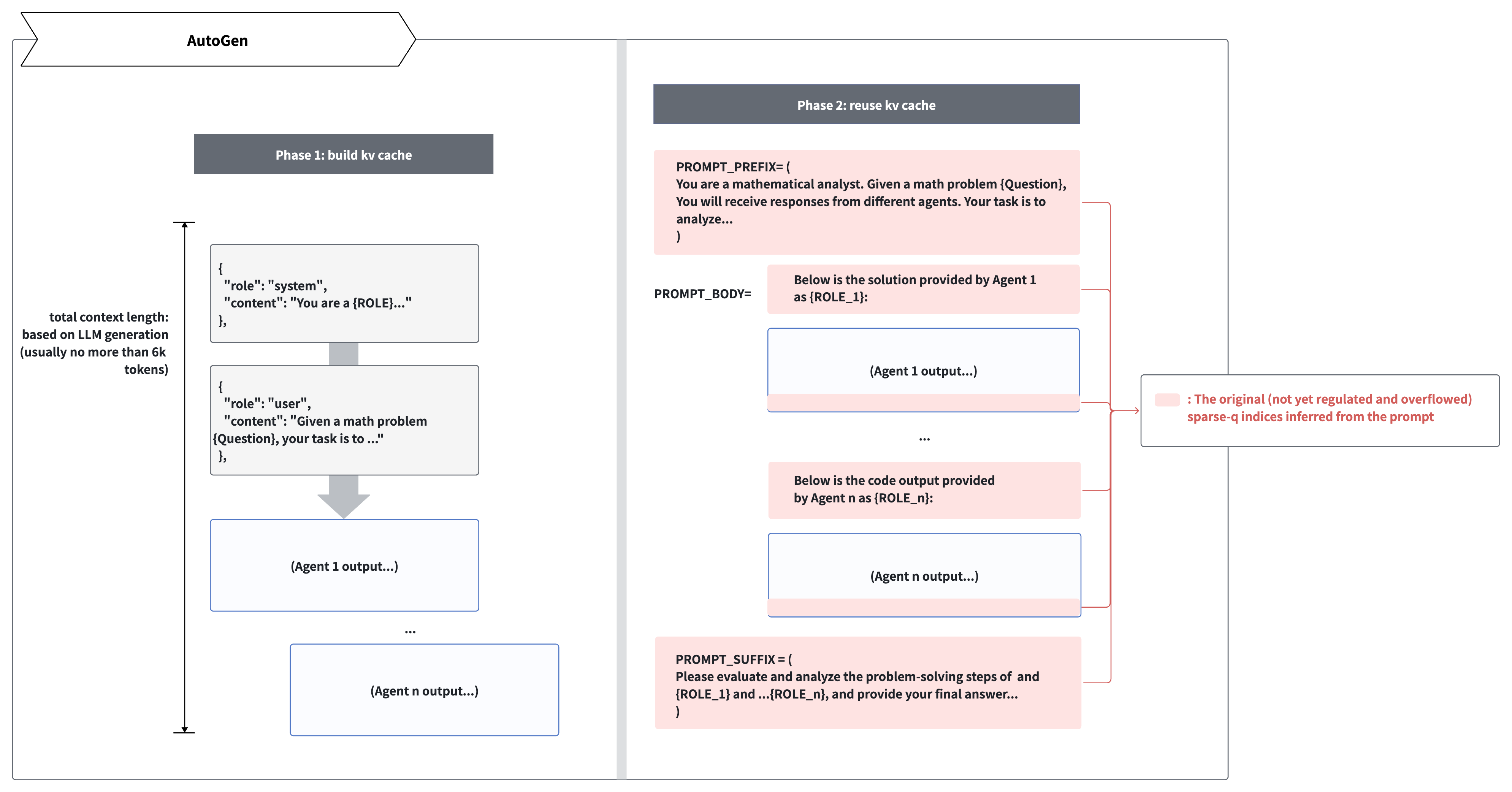}
    \caption{Prompt construction for AutoGen-style agent workflows.}
    \label{fig:app-autogen}
\end{figure}

\subsection{Multi-Agent Debate}

For the multi-agent debate setting, the total context length is also determined by LLM generation and is usually no more than 8K tokens. Phase 1 builds reusable \kv{} blocks from the debaters' messages and generated arguments. In Phase 2, the prefix defines the moderator role and the debate topic. The prompt body reuses the affirmative and negative arguments together with the corresponding agent outputs. The suffix asks the moderator to evaluate both sides and determine whether there is a clear preference for an answer candidate.

\begin{figure}[H]
    \centering
    \includegraphics[width=\linewidth]{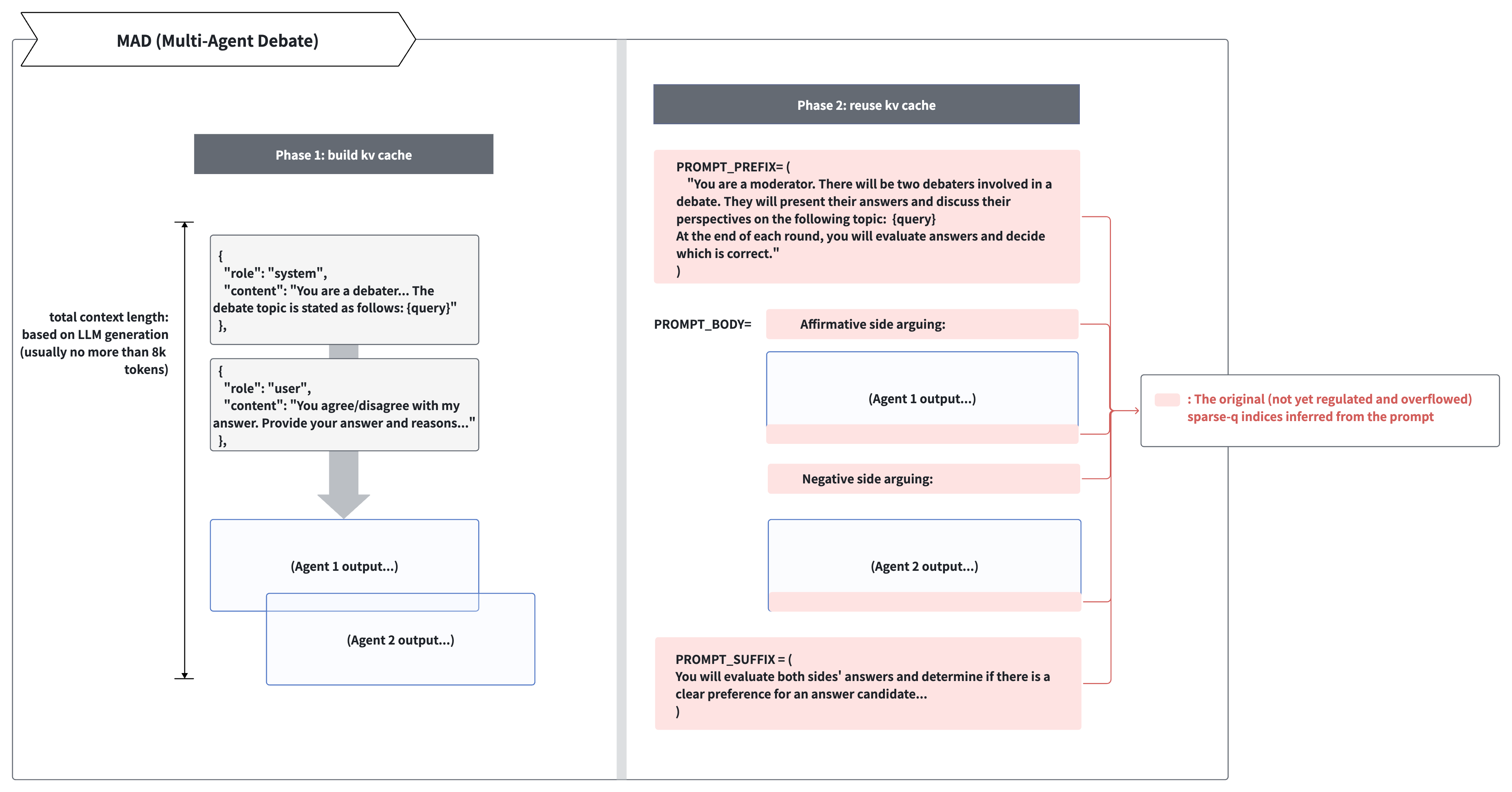}
    \caption{Prompt construction for multi-agent debate.}
    \label{fig:app-mad}
\end{figure}

\end{document}